\documentclass[final,5p,twocolumn]{elsarticle}

\ifpreprint \usepackage[left=1.5cm,right=1.5cm]{geometry} \fi

\biboptions{sort&compress}

\usepackage{amssymb}

\usepackage{amsmath}

\usepackage{pgfplots}

\usepackage[hidelinks]{hyperref}

\usepackage{mathtools}

\usepackage{graphicx}

\usepackage{microtype}

\usepackage{float}

\usepackage{booktabs}
\usepackage[bitstream-charter]{mathdesign}

\usepackage{dblfloatfix}

\usepackage{xcolor}
\definecolor{TUMBlue}{HTML}{0065bd}
\definecolor{TUMRed}{HTML}{e37222}
\definecolor{TUMGreen}{HTML}{a2ad00}
\definecolor{TUMPink}{HTML}{b55ca5}

\usepackage{ulem}

\journal{Journal of Energy Storage}

\begin{document}

\begin{frontmatter}

\title{Time-dependent global sensitivity analysis of the Doyle-Fuller-Newman model}

\author[1]{Elia Zonta\corref{cor1}}
\ead{elia.zonta@tum.de}
\author[2]{Ivana Jovanovic Buha}
\author[3]{Michele Spinola}
\author[3]{Christoph Weißinger}
\author[2]{Hans-Joachim Bungartz}
\author[1]{Andreas Jossen}

\cortext[cor1]{Corresponding author}

\affiliation[1]{organization={Technical University of Munich, School of Engineering and Design, Department of Energy and Process Engineering, Chair of Electrical Energy Storage Technology}, 
            addressline={Arcisstraße 21},
            city={Munich},
            postcode={80333},
            country={Germany}}

\affiliation[2]{organization={Technical University of Munich, School of Computation, Information and Technology, Department of Computer Science, Chair of Scientific Computing}, 
            addressline={Boltzmannstraße 3},
            city={Garching},
            postcode={85748},
            country={Germany}}

\affiliation[3]{organization={Capgemini Engineering, Center of Excellence Battery}, 
            addressline={Frankfurter Ring 81}, 
            city={Munich},
            postcode={80807},
            country={Germany}}

\begin{abstract}
The Doyle-Fuller-Newman model is arguably the most ubiquitous electrochemical model in lithium-ion battery research. Since it is a highly nonlinear model, its input-output relations are still poorly understood. Researchers therefore often employ sensitivity analyses to elucidate relative parametric importance for certain use cases. However, some methods are ill-suited for the complexity of the model and appropriate methods often face the downside of only being applicable to scalar quantities of interest. We implement a novel framework for global sensitivity analysis of time-dependent model outputs and apply it to a drive cycle simulation. We conduct a full and a subgroup sensitivity analysis to resolve lowly sensitive parameters and explore the model error when unimportant parameters are set to arbitrary values. Our findings suggest that the method identifies insensitive parameters whose variations cause only small deviations in the voltage response of the model. By providing the methodology, we hope research questions related to parametric sensitivity for time-dependent quantities of interest, such as voltage responses, can be addressed more easily and adequately in simulative battery research and beyond.
\end{abstract}




\end{frontmatter}

\section{Introduction}
Necessitated by the ongoing efforts to reduce carbon emissions and thereby mitigate the effects of climate change, lithium-ion batteries have become increasingly popular in the electrification of transport and the storage of renewable energy~\cite{chu_opportunities_2012}. The intensification of battery research focused on improved energy density, lifetime, and optimal utilization therefore lies at the forefront of attenuating the impact of global warming by enabling a renewable energy transition and reducing greenhouse gas emissions.

Computational models are indispensable to many of the aforementioned research areas. Whether it be fast charging protocols~\cite{oehler_online_2024}, battery pack design for automotive applications~\cite{astaneh_multiphysics_2022}, or the elucidation of dominant aging mechanisms~\cite{zhang_novel_2020}, simulations enable the cost-efficient investigation of systems of interest, the estimation of internal quantities \textit{in operando} and facilitate scientists to vary experimentally inaccessible control variables in numerical experiments.

\begin{figure}[pbht]
\centering
\includegraphics[width=0.5\textwidth]{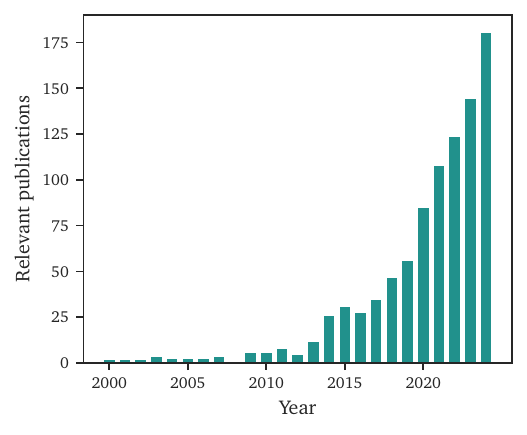}
\caption{The number of publications on Scopus mentioning "lithium-ion battery" and "sensitivity analysis" in either title, abstract, or keywords over the past years, indicating a clear upwards trend.}\label{fig1}
\end{figure}

However, the employed models in battery research tend to be very complex and comprise a large number of parameters. Arguably the most prominent electrochemical model for battery dynamics is the Doyle-Fuller-Newman (DFN) or pseudo-two-dimensional (p2D) model, conceived by the eponymous Doyle, Fuller, and Newman~\cite{doyle_modeling_1993,fuller_simulation_1994}. It is composed of partial differential-algebraic equations describing charge transport, mass transport, and electrochemical reactions. Researchers therefore often seek to reduce the complexity of these models and alleviate their parametrization by only considering the most influential model parameters. This has led to a substantial increase in conducted sensitivity analyses in battery research (Fig.~\ref{fig1}). Sensitivity analysis aims at quantifying the impact of a parameter variation on the model output in the form of a sensitivity index, in order to obtain a relative importance ranking of the model parameters.

There exist various approaches to perform sensitivity analysis, but not all are equally suited for the evaluation of complex models. In fact, Saltelli et al.~\cite{saltelli_why_2019, saltelli_how_2010} highlight misconceptions in a considerable number of published sensitivity analyses across numerous scientific disciplines. One of their biggest concerns is the application of one-at-a-time (OAT) methods to highly nonlinear models, which restricts the validity of the obtained results and hinders straightforward interpretation. In battery research, OAT methods are arguably the most prevalent practice for determining parametric sensitivity~\cite{edouard_parameter_2016, bi_automatic_2018, li_parameter_2020, liu_simulation_2020, vazquez-arenas_rapid_2014, zhang_parameter_2014, wang_parameter_2024, li_data-driven_2022, pan_parameter_2022, wimarshana_parameter_2022, lai_parameter_2020, chang_electrochemical_2024, rabissi_fast_2023}. In light of the nonlinearity of the DFN model and similar battery models, this is a widely made but nonetheless suboptimal choice.

The use of global sensitivity analysis, i.e., methods suitable for nonlinear models, is comparatively scarce. Among the available methods, variance-based sensitivity analysis is probably the most commonly employed technique, besides the elementary effect test~\cite{xu_enabling_2022,zhao_enhancing_2024}, also known as Morris method. Scheller et al.~\cite{scheller_impact_2024} use variance-based sensitivity analysis and a finite element all-solid-state battery model to determine key parameters critical for cell performance and their related optimization potential. Streb et al.~\cite{streb_improving_2022} utilize variance-based sensitivity analysis to devise optimal experimental designs for the data-driven parametrization of a DFN model. Lin et al.~\cite{lin_efficient_2018} implement a framework for efficient variance-based sensitivity analysis and apply it to a three-dimensional lithium-ion battery model. Rüther et al.~\cite{ruther_demystifying_2024} use variance-based sensitivity analysis to investigate a numerical impedance model. Ko et al.~\cite{ko_using_2024} employ variance-based sensitivity analysis to determine the most relevant parameters serving as battery aging indicators, with subsequent parameter identification at different states of health. Chen et al.~\cite{chen_global_2025} use variance-based sensitivity analysis to find operating conditions that increase the sensitivity of specific parameters in an electrochemical-thermal battery model. While global sensitivity analysis is adequate for nonlinear models, the main drawback of the above works is that commonly available global sensitivity analysis approaches can only deal with scalar-valued quantities of interest. For time-dependent quantities, like the terminal voltage of a battery during charging or discharging, it is therefore necessary to resort to integrals over time, averages, or the selection of a single point in time. This is disadvantageous because integrating or averaging the quantity of interest over time might lead to a loss of information, and only selecting one or multiple points in time does not yield a comprehensive picture.
 
In order to alleviate the issue of arbitrarily producing a scalar value from a vectorial simulation result for sensitivity analysis, this work therefore implements an innovative framework for global sensitivity analysis of time-dependent quantities of interest and applies it to the voltage response of the DFN model for a drive cycle with high dynamics. To the best of our knowledge, this is the first time parametric sensitivity is determined over a dynamic battery model output in a statistically rigorous fashion. This marks a significant leap in methodology and can aid future simulative battery research in dealing with research questions related to parametric sensitivity more robustly. Our particular contributions are the following:
\begin{itemize}
    \item We reiterate the arguments from literature calling for alternatives to the OAT method in the case of nonlinear battery models (Section~\ref{sec:inadequacy})
    \item We summarize the theoretical background of the novel sensitivity analysis methodology devised by Alexanderian et al.~\cite{alexanderian_2020} (Section~\ref{sec:alexanderian}) and provide the open-source implementation used in this work.
    \item By leveraging the \texttt{LiionDB}~\cite{wang_review_2022}, we show the wide spread of battery model parameters found in the literature, even in the case of identical electrode materials, which hinders parametrization from literature-derived values (Fig.~\ref{fig:param_range}).
    \item Using the aforementioned framework, we conduct a time-dependent global sensitivity analysis of the DFN model for a drive cycle (Section~\ref{sec:full}) and a subgroup analysis (Section~\ref{sec:subgroup}) over the previously determined parameter range. In contrast to previous works, our study aggregates the parametric effects over the whole time-dependent trajectory and allows for a separation of first- and total-order effects.
    \item Furthermore, we investigate the model error when setting an increasing number of insensitive parameters to arbitrary values from literature (Section~\ref{sec:error}) as a guidance for battery modelling practitioners, given the widespread reliance on literature-based parameter values.
\end{itemize}

\section{Background}
\label{sec:background}

\subsection{Sensitivity analysis}
The field of sensitivity analysis deals with the question of how to obtain a relative importance ranking of the parameters of a mathematical model. In this context, simulators are often seen as black-box functions of the general form
\begin{equation}
    Y = f(\boldsymbol{\xi}), \quad  \boldsymbol{\xi} \in \mathbb{R}^d,
\end{equation}
where $\boldsymbol{\xi}$ is a $d$-dimensional vector of uncertain model parameters and $Y$ is the corresponding model output. Elucidating the relationship between $\boldsymbol{\xi}$ and $Y$ can be very challenging when $f$ is a complex model.

Methods for sensitivity analysis are categorized into local and global approaches~\cite{razavi_what_2015}. While local sensitivity analysis investigates the sensitivity around a single point in parameter space, global sensitivity analysis aims at providing a full picture across the whole parameter space.

\subsubsection{The inadequacy of one-at-a-time methods} \label{sec:inadequacy}
The OAT method is based on the idea of coarsely approximating the gradient of the response surface of a model around a reference point in parameter space in a finite difference fashion~\cite{razavi_what_2015}. This is generally achieved by spanning a parameter space with minimum and maximum values for each considered parameter and then varying each parameter separately around its nominal value. By design, this approach cannot account for interactions of parameter combinations, since multiple parameters are never varied concomitantly. 

\begin{figure}[htbp]
    \centering
    \includegraphics[width=0.5\textwidth]{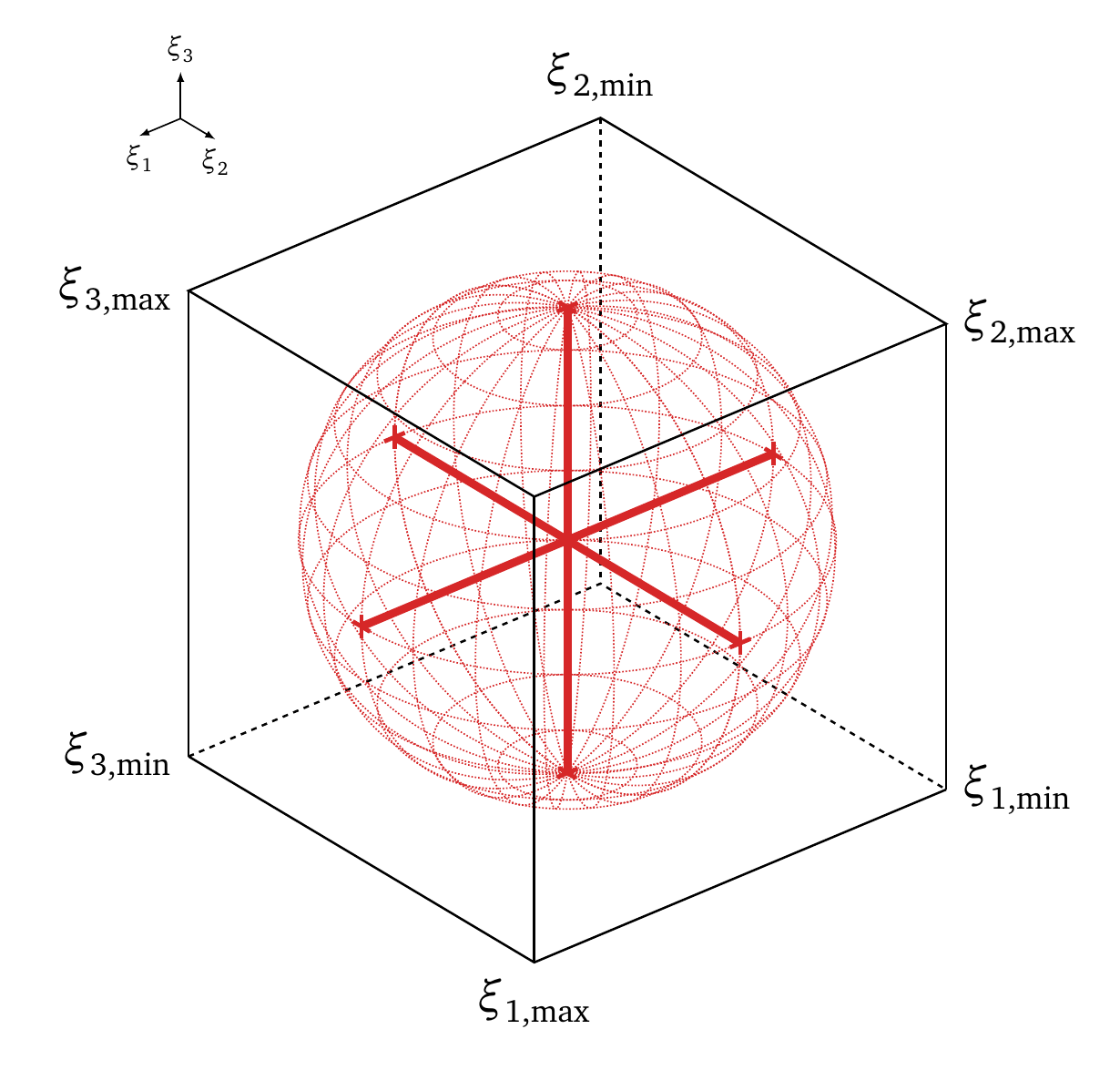} 
    \caption{Visualization of a parameter space spanned by the three parameters $\xi_1$, $\xi_2$, and $\xi_3$. The hypercross structure sampled by OAT methods and its enveloping hypersphere inhabit only a fraction of the entire volume. This fraction decreases very rapidly as the number of parameters increases~\cite{saltelli_why_2019}.}
    \label{fig:OAT-parameter-space}
\end{figure}

The much bigger concern according to Saltelli et al.~\cite{saltelli_why_2019}, however, is that the OAT method leaves the defined parameter space vastly underexplored. Varying parameters individually around their nominal value limits the exploration of the parameter space to a ``hypercross'' structure. In high-dimensional parameter spaces, this hypercross has neglectable coverage when compared to the entire volume of the space. This problem is illustrated well by imagining the enveloping hypersphere of the hypercross (Fig.~\ref{fig:OAT-parameter-space}) and computing volume ratios between the sphere and the entire parameter space. In 10 dimensions, one would obtain a volume ratio of $2.5\cdot 10^{-3}$, which is a very generous upper bound estimate considering the hypercross constitutes only a very small fraction of the enveloping hypersphere. Therefore, the OAT approach leaves most of the parameter space unexplored for higher dimensional problems. For the sake of clarity, we complement these theoretical considerations with an illustrative example from literature in \ref{app3}.

Due to these limitations, OAT methods are considered unsuitable for nonlinear models~\cite{saltelli_why_2019, saltelli_how_2010, czitrom_one-factor-at--time_1999}. Their simplicity and ease of use make them a widely employed choice regardless, especially in battery research.

\subsubsection{Variance-based sensitivity analysis}
Variance-based sensitivity analysis tries to apportion the variance of the model response to each individual parameter, thereby providing a measure of the parameter's importance. This is facilitated by the Hoeffding-Sobol' decomposition of $f$, which is a sum of functions of increasing dimension~\cite{hoeffding_class_1948, sobol_global_2001}
\begin{equation} \label{eq:hoeffding-sobol}
    f(\boldsymbol{\xi}) = f_0 + \sum_{i=1}^d f_i(\xi_i) + \sum_{i<j}^d f_{ij}(\xi_j, \xi_j) + ... + f_{1,2,...,d}(\xi_1, \xi_2, ..., \xi_d),
\end{equation}
where $f_0$ is a constant, the $f_i$ are one-dimensional functions depending only on the $i$-th component of $\boldsymbol{\xi}$, $f_{i,j}$ are two-dimensional functions depending on the $i$-th and $j$-th component, etc. If we now consider $\boldsymbol{\xi}$ a random vector
\begin{equation}
    \boldsymbol{\xi} = \{\xi_1,..., \xi_d\}, \quad \text{with} \quad \xi_i\sim\mathcal{P}_i, \quad i\in\{1,...,d\}, 
\end{equation}
where each component $\xi_i$ is a random variable distributed according to its probability distribution $\mathcal{P}_i$, the model output $Y=f(\boldsymbol{\xi})$ becomes a random variable~\cite{sudret_global_2008}. Using (\ref{eq:hoeffding-sobol}), the total variance of the output $D=\mathbb{V}[Y]$ can similarly be decomposed into partial variances~\cite{saltelli_variance_2010} 
\begin{equation} \label{eq:variance_decomposition}
    D = \sum_i^d D_i + \sum_{i<j}^d D_{ij} + ... + D_{1,2,...,d},
\end{equation}
which, upon dividing by $D$, yields the Sobol' indices of all orders
\begin{equation}
    1 = \sum_i^d S_i + \sum_{i<j}^d S_{ij} + ... + S_{1,2,...,d}. 
\end{equation}
The $S_i$ are called \textit{first-order} or \textit{main effect} Sobol' indices, as they relate the variance caused solely by individual parameters to the total variance. The \textit{total-order} or \textit{total effect} Sobol' indices $S_{i}^{\text{tot}}$ represent the effect of a parameter when taking all possible higher-order interactions with other parameters into account. They are defined as~\cite{sudret_global_2008}
\begin{equation} \label{eq:total_sobol}
    S_{i}^{\text{tot}} = \frac{\sum_{k\in\mathcal{K}_i} D_k}{D} =  \frac{D_i + D_{ij} + ... + D_{1,...,i,...,d}}{D},
\end{equation}
where $\mathcal{K}_i$ is an index set containing all combinations that include $i$. The expression in (\ref{eq:total_sobol}) is equivalent to the sum of all Sobol' indices that consider the $i$-th parameter. Sobol' indices therefore provide a sensitivity measure by quantifying the proportion of the total uncertainty that is caused by each individual input parameter of the model.

\subsubsection{Extension to time-dependent quantities of interest}
\label{sec:alexanderian}

In the above derivation, the model response $Y$ was considered a scalar. Alexanderian et al.~\cite{alexanderian_2020} introduced novel approaches for conducting variance-based sensitivity analyses for time-dependent model outputs, which will be summarized here. In the following, we will consider a time-dependent model
\begin{equation}
    Y = f(t,\boldsymbol{\xi}), \quad t \in [0,T].
\end{equation}
While it was previously possible to compute Sobol' indices or other global sensitivity metrics for each point in time $t_k$ of a grid
\begin{equation}
    \{t_k\}_{k=1}^{N_{\text{quad}}} \quad \text{with} \quad t_{k-1} < t_k \quad \text{and} \quad t_1 = 0, \quad t_{N_{\text{quad}}}=T,
\end{equation}
as done by, e.g., Hadigol et al.~\cite{hadigol_uncertainty_2015} or Constantine and Doostan~\cite{constantine_timedependent_2017}, this approach presents two major shortcomings. Firstly, since models in science and engineering are generally heteroscedastic, i.e., their variance is not constant over time, the relative importance of parameters over time is inconclusive, as parameters contributing heavily to a comparably small variance should be overall considered less important. Secondly, the temporal correlation structure of the model response is neglected, because each $f(t_k,\boldsymbol{\xi})$ is treated independently.

For the aforementioned reasons, Alexanderian et al.~\cite{alexanderian_2020} devised algorithms based on surrogate models and spectral representations that alleviate both issues and will be outlined in the following. The underlying idea is the generalization of the previously defined classical Sobol' indices
\begin{equation}
    S_i = \frac{D_i}{D}, \quad S_i^{\text{tot}} = \frac{\sum_{k\in\mathcal{K}_i} D_k}{D}
\end{equation}
to time-dependent variances
\begin{equation} \label{eq:sobol-time-dependent}
    \mathfrak{S}_i(T) = \frac{\int_0^T D_i(t) \, dt}{\int_0^T D(t) \, dt}, \quad \mathfrak{S}_i^{\text{tot}}(T) = \frac{\int_0^T \sum_{k\in\mathcal{K}_i} D_k(t) \, dt}{\int_0^T D(t) \, dt}.
\end{equation}
By employing a quadrature rule with nodes and weights $\{t_m, w_m\}_{m=1}^{N_{\text{quad}}}$ on the interval $[0,T]$, we can solve the integrals in (\ref{eq:sobol-time-dependent}) numerically via
\begin{equation} \label{eq:sobol-quadrature}
    \begin{split}
        \mathfrak{S}_i(T) &\approx \frac{\sum_{m=1}^{N_{\text{quad}}} w_m D_i(t_m)}{\sum_{m=1}^{N_{\text{quad}}} w_m D(t_m)}, \\ 
        \quad \mathfrak{S}_i^{\text{tot}}(T) &\approx \frac{\sum_{m=1}^{N_{\text{quad}}} \sum_{k\in\mathcal{K}_i} w_m D_k(t_m)}{\sum_{m=1}^{N_{\text{quad}}} w_m D(t_m)}.
    \end{split}
\end{equation}
By constructing pointwise-in-time polynomial chaos (PC) surrogate models, one can obtain the partial variances $D_i(t_m)$ and $D_k(t_m)$. A PC expansion is generally of the form
\begin{equation}
    f(t,\boldsymbol{\xi}) = \sum_{k=1}^{\infty} c_k(t) \Psi_k(\boldsymbol{\xi}) \approx \sum_{k=1}^{N_{\text{PC}}} c_k(t) \Psi_k(\boldsymbol{\xi}),
\end{equation}
where $\{\Psi_k\}_{k=1}^{N_{\text{PC}}}$ is a set of orthogonal polynomials or an orthonormal polynomial basis, the $\{c_k\}_{k=1}^{N_{\text{PC}}}$ are expansion coefficients and $N_{\text{PC}}$ is the finite number of terms the exact PC expansion is truncated to. Since $\boldsymbol{\xi}$ is a $d$-dimensional random vector, the $\Psi_k$ are $d$-variate polynomials. The benefit of PC expansion-based surrogate models lies in their ability to provide analytical expressions for the partial variances~\cite{sudret_global_2008}. This enables us to rewrite (\ref{eq:sobol-quadrature}) as
\begin{equation} \label{eq:PC-sobol}
    \begin{split}
        \mathfrak{S}_i(T) &\approx \frac{\sum_{m=1}^{N_{\text{quad}}} \sum_{k\in\mathcal{J}_i} ||\Psi_k||^2 w_m c_k(t_m)^2}{\sum_{m=1}^{N_{\text{quad}}} \sum_{k=1}^{N_{\text{PC}}} ||\Psi_k||^2 w_m c_k(t_m)^2}, \\
        \mathfrak{S}_i^{\text{tot}}(T) &\approx \frac{\sum_{m=1}^{N_{\text{quad}}} \sum_{k\in\mathcal{K}_i} ||\Psi_k||^2 w_m c_k(t_m)^2}{\sum_{m=1}^{N_{\text{quad}}} \sum_{k=1}^{N_{\text{PC}}} ||\Psi_k||^2 w_m c_k(t_m)^2},
    \end{split}
\end{equation}
where $\mathcal{J}_i$ is an index set that includes all terms in the PC expansion that consider only the $i$-th parameter. If the $\Psi_k$ form an orthonormal basis, then $||\Psi_k||^2=1\,\forall k$.

The second approach from Alexanderian et al.~\cite{alexanderian_2020} for computing the generalized Sobol' indices relies on spectral representations, namely Karhunen-Loève (KL) expansions. The KL expansions are of the form
\begin{equation}
    f(t,\boldsymbol{\xi}) = f_0 + \sum_{i=1}^{\infty} f_i(\boldsymbol{\xi})e_i(t) \approx f_0 + \sum_{i=1}^{N_{\text{KL}}} f_i(\boldsymbol{\xi})e_i(t),
\end{equation}
with $f_0$ being the mean of the time-dependent process $\mathbb{E}\left[f(t,\boldsymbol{\xi})\right]$, the $f_i(\boldsymbol{\xi})$'s are expansion coefficients with variance $\mathbb{V}\left[f_i(\boldsymbol{\xi})\right]=\lambda_i$, and $\lambda_i$ and $e_i$ are the eigenvalues and eigenvectors of the covariance operator of $f(t,\boldsymbol{\xi})$.

To construct a KL expansion, $f(t,\boldsymbol{\xi})$ needs to be a centered process, i.e., it needs to have zero mean. The first step is therefore to center the process by subtracting the mean from all $N$ model evaluations $\{f(t_m, \boldsymbol{\xi}^j)\}_{j=1}^N$ at each point in time $t_m$
\begin{equation}
    f_c(t_m,\boldsymbol{\xi}^k) = f(t_m, \boldsymbol{\xi}^k) - \frac{1}{N} \sum_{j=1}^N f(t_m,\boldsymbol{\xi}^j),
\end{equation}
with $k\in\{1,...,N\}$ and $m\in\{1,...,N_{\text{quad}}\}$.

Then, we can construct the covariance matrix $\boldsymbol{K}$ of the centered function evaluations by obtaining its elements $K_{lm}$ via
\begin{equation}
    K_{lm} = \frac{1}{N-1} \sum_{k=1}^N f_c(t_l,\boldsymbol{\xi}^k) f_c(t_m,\boldsymbol{\xi}^k),
\end{equation}
with $l,m\in\{1,...,N_{\text{quad}}\}$.

The solution of the eigenvalue problem
\begin{equation}
    \begin{split}
    \boldsymbol{W}^{\frac{1}{2}} \boldsymbol{K} \boldsymbol{W}^{\frac{1}{2}} \boldsymbol{u}_i = \lambda_i \mathbf{u}_i, \\
    \text{with} \quad \boldsymbol{W}=\text{diag}(w_1,w_2,...,w_{N_{\text{quad}}}),\quad& i\in\{1,...,N_{\text{quad}}\},
    \end{split}
\end{equation}
specifically the spectral decay of the eigenvalues $\lambda_i$ , which informs the truncation level $N_{\text{KL}}$, lies at the heart of this approach. The truncation level should be chosen such that the fraction of the total variance captured by the truncated KL expansion, given as
\begin{equation}
    r_{N_{\text{KL}}} = \frac{\sum_{i=1}^{N_{\text{KL}}}\lambda_i}{\sum_{i=1}^{\infty} \lambda_i},
\end{equation}
is sufficiently large. If one can obtain a large fraction $r_{N_{\text{KL}}}$ with a small truncation level $N_{\text{KL}}$, the respective process is deemed a \textit{low-rank} process. This means that almost all the uncertainty in $f$ is quantified by the $N_{\text{KL}}$ KL modes.

\begin{figure}[hpbt]
    \centering
    \includegraphics[width=0.5\textwidth]{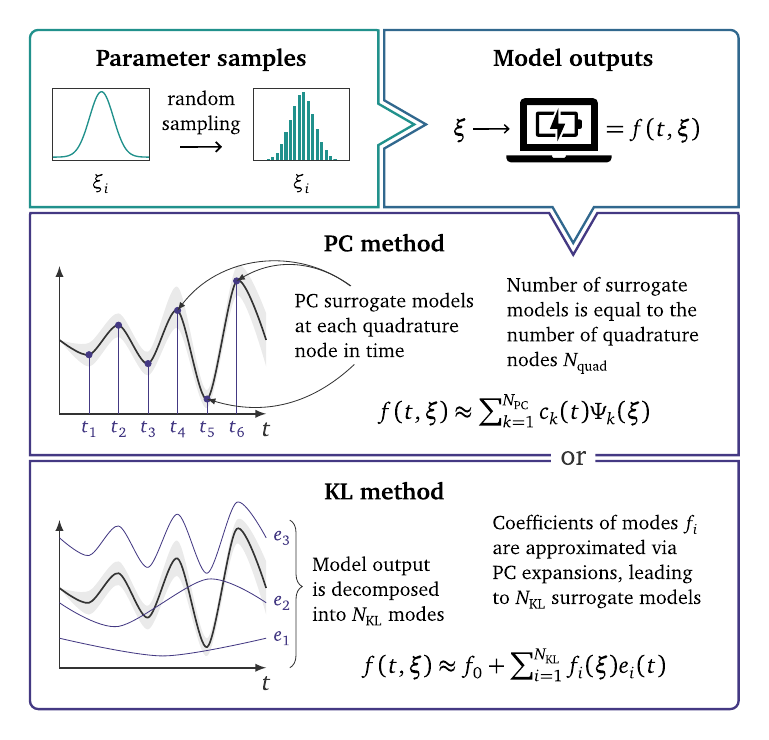}
    \caption{A visual summary of the main differences between the PC method and the KL method.}
    \label{fig:overview_PC_KL}
\end{figure}

To compute the KL modes, we first need to obtain the eigenvectors $\boldsymbol{e}_i$ from $\boldsymbol{u}_i$ via
\begin{equation}
    \boldsymbol{e}_i = \boldsymbol{W}^{-\frac{1}{2}} \boldsymbol{u}_i, \quad i\in\{1,...,N_{\text{quad}}\}.
\end{equation}
We can then compute 
\begin{equation}
    f_i(\boldsymbol{\xi^k}) = \sum_{m=1}^{N_{\text{quad}}} w_m f_c(t_m, \boldsymbol{\xi^k}) e_i^m,
\end{equation}
with $i\in\{1,...,N_{\text{KL}}\}$ and $k\in\{1,...,N\}$. This leaves us with $N$ realizations $f_i(\boldsymbol{\xi}^k)$ of each mode $f_i$, which we can use to construct a PC surrogate model for each mode
\begin{equation}
    f_i(\boldsymbol{\xi}) \approx \sum_{k=1}^{N_{\text{PC}}} c_k^i \Psi_k(\boldsymbol{\xi}).
\end{equation}
Using these PC expansions, we can again efficiently compute the generalized Sobol' indices as
\begin{equation} \label{eq:KL-sobol}
    \begin{split}
        \mathfrak{S}_i(T) &\approx \frac{\sum_{i=1}^{N_{\text{KL}}} \sum_{k\in\mathcal{J}_i} ||\Psi_k||^2 (c_k^i)^2}{\sum_{k=1}^{N_{\text{KL}}} \lambda_i}, \\
        \mathfrak{S}_i^{\text{tot}}(T) &\approx \frac{\sum_{i=1}^{N_{\text{KL}}} \sum_{k\in\mathcal{K}_i} ||\Psi_k||^2 (c_k^i)^2}{\sum_{k=1}^{N_{\text{KL}}} \lambda_i}.
    \end{split}
\end{equation}
In the following, we will refer to the approaches defined by equation (\ref{eq:PC-sobol}) and equation (\ref{eq:KL-sobol}) as PC method and KL method respectively. Please note that while both methods employ PC expansions, the PC method explicitly refers to the use of pointwise-in-time PC surrogate models. Fig.~\ref{fig:overview_PC_KL} shortly summarizes the main differences between the two methods. It should be noted that for the analysis, all model outputs need to span the same time interval, which limits its applicability in cases where the simulated duration varies between runs or depends on the input parameters. This could be the case in (dis-)charge simulations with a constant current, when the sampled parameters lead to varying battery capacities.

\subsection{The Doyle-Fuller-Newman model}
\label{sec:DFN}
\begin{figure}[tbph]
    \centering
    \includegraphics[width=0.5\textwidth]{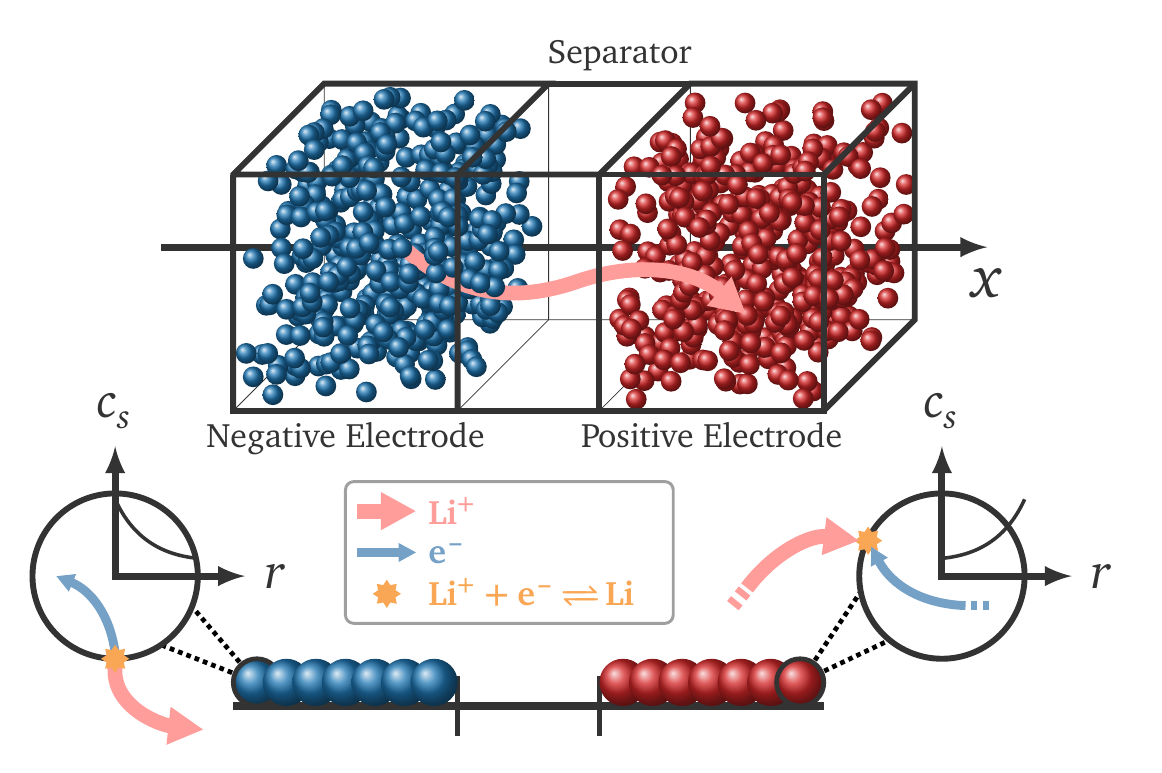}
    \caption{Schematic depiction of the DFN model during discharge.}
    \label{fig:DFN}
\end{figure}

The DFN or p2D model was devised by Doyle, Fuller, and Newman~\cite{doyle_modeling_1993, fuller_simulation_1994} in the early 1990s and has since become an invaluable tool in battery research. It is based on porous electrode theory and Butler-Volmer kinetics to describe the relevant battery processes on the continuum scale. To this end, the cell model is divided into a positive electrode, separator, and negative electrode domain. Each of these domains is thought of as a porous medium filled with electrolyte, and the particles in the electrode domains are considered perfectly spherical. This porous medium, however, is not fully resolved but rather homogenized and projected onto the $x$-dimension. The spherical particles are placed along this one-dimensional axis in the positive and negative electrode domains to act as representative electrode particles (Fig.~\ref{fig:DFN}). The diffusion of lithium inside the solid part of the electrode is resolved over the particle's radius.

The constituting equations of the DFN model can be split into three distinct groups, namely equations that describe charge conservation, mass conservation, and electrochemical kinetics. In the following, we closely adhere to the conventions of Sulzer et al.~\cite{sulzer_python_2021, pybammDocumentation}. Since the model describes a lithium-ion cell, the number of transferred electrons is $z=1$ and all involved ions are singly charged.

\textit{Charge conservation:}
The spatial gradient of the current in the electrolyte $i_{\text{l}}$ is zero in the separator, but equal to the intercalation current in the electrode domains
\begin{equation}
    \frac{\partial i_{\text{l,k}}}{\partial x} = \begin{cases}
        a_{\text{k}}j_{\text{k}}, & \mathrm{k} \in \{\mathrm{neg,\,pos}\} \\
        0, & \mathrm{k = sep},
    \end{cases}
\end{equation}
where $a$ is the specific surface area and $j$ the intercalation current density. The index $k\in\{\mathrm{neg,\,sep,\,pos}\}$ describes the domain on which the equations hold, with $\text{neg}$ denoting the negative electrode domain, $\text{sep}$ the separator domain, and $\text{pos}$ the positive electrode domain. The current in the electrolyte itself can be obtained via
\begin{equation}
    \begin{split}
        i_{\text{l,k}} = \varepsilon_{\text{k}}^{b_{\text{k}}}\kappa \left( -\frac{\partial \phi_{\text{l,k}}}{\partial x} + 2\Theta(1-t^+) \frac{RT}{F} \frac{\partial \ln c_{{\text{l,k}}}}{\partial x} \right), \\ \mathrm{k} \in \{\mathrm{neg,\,sep,\,pos}\},
    \end{split}
\end{equation}
where $\varepsilon$ is the porosity, $b$ the Bruggeman coefficient, $\phi_\text{l}$ is the potential in the liquid phase, $t^+$ is the transference number of the lithium ions, $R$ is the universal gas constant, $F$ the Faraday constant, $T$ is the temperature, $c_\text{l}$ is the concentration of lithium ions in the electrolyte, and $\Theta \coloneqq 1 + \frac{\partial \ln f}{\partial \ln c_{{\text{l}}}}$ is the so-called \textit{thermodynamic factor}, which is treated as a parameter. It also holds that
\begin{equation}
    I - i_{{\text{l,k}}} = -\sigma_\text{k} \frac{\partial \phi_{{\text{s,k}}}}{\partial x}, \quad \mathrm{k} \in \{\mathrm{neg,\,pos}\},
\end{equation}
where $I$ is the total applied current density, $\sigma$ is the electric conductivity, and $\phi_\text{s}$ is the potential in the solid phase. Since there is no current flowing through the solid phase in the separator, we have $I- i_{{\text{l,sep}}} = 0$.

\textit{Mass conservation:}
The change in lithium ion concentration in the electrolyte $c_\text{l}$ over time is given by
\begin{equation}
    \varepsilon_\text{k} \frac{\partial c_{{\text{l,k}}}}{\partial t} = - \frac{\partial N_{\text{l,k}}}{\partial x} + \frac{1}{F} \frac{\partial i_{\text{l,k}}}{\partial x}, \quad \mathrm{k} \in \{\mathrm{neg,\,sep,\,pos}\}
\end{equation}
with the molar flux in the liquid phase $N_\text{l}$ defined as
\begin{equation}
    N_{\text{l,k}} = -\varepsilon_\text{k}^{b_\text{k}} D_\text{l} \frac{\partial c_{\text{l,k}}}{\partial x} + \frac{t^+}{F} i_{\text{l,k}}, \quad \mathrm{k} \in \{\mathrm{neg,\,sep,\,pos}\},
\end{equation}
where $D_\text{l}$ is the diffusion coefficient of the liquid phase. Lithium ion transport in the solid phase is described by Fickian diffusion
\begin{equation}
    \frac{\partial c_{\text{s,k}}}{\partial t} = - \frac{1}{r^2_\text{k}} \frac{\partial}{\partial r_\text{k}} (r_\text{k}^2 N_{\text{s,k}}), \quad \mathrm{k} \in \{\mathrm{neg,\,pos}\},
\end{equation}
where $c_\text{s}$ is the concentration of lithium in the solid phase, $r$ denotes the radial coordinate of the spherical electrode particle, and the molar flux in the solid phase $N_\text{s}$ is
\begin{equation}
    N_{\text{s,k}} = - D_{\text{s,k}} \frac{\partial c_{\text{s,k}}}{\partial r_\text{k}}, \quad \mathrm{k} \in \{\mathrm{neg,\,pos}\},
\end{equation}
with $D_\text{s}$ being the diffusion coefficient of the solid phase.

\textit{Electrochemical kinetics:}
The electrode current density $j$ is given by the Butler-Volmer equation
\begin{equation}
    \begin{split}
    j_\text{k} = j_{0,\text{k}} \left( \exp \left(  \frac{F\alpha_\text{ox}}{RT} \eta_\text{k} \right) - \exp \left(  -\frac{F\alpha_\text{red}}{RT} \eta_\text{k} \right) \right), \\ \mathrm{k} \in \{\mathrm{neg,\,pos}\},
    \end{split}
\end{equation}
where $\alpha_\text{ox}=\alpha_\text{red}=0.5$ are the charge transfer coefficients of the oxidation and reduction reaction respectively, $\eta$ stands for the overpotential, and $j_0$ is the exchange current density, computed as
\begin{equation}
\label{eq:j0}
    j_{0,\text{k}} = F k_{0,\text{k}} c_{\text{s,k}}^{\alpha_\text{red}} (1 - c_{\text{s,k}})^{\alpha_\text{ox}} c_{\text{l,k}}^{\alpha_\text{ox}}\bigr\rvert_{r_\text{k}=R_\text{k}}, \quad \mathrm{k} \in \{\mathrm{neg,\,pos}\}, 
\end{equation}
with $k_0$ as the reaction rate constant and $R_\text{k}$ being the particle radius. Note that different conventions exist regarding the form of equation (\ref{eq:j0}), warranting careful consideration of the units of the reaction rate constant. The overpotential is defined as
\begin{equation}
    \eta_\text{k} = \phi_{\text{s,k}} - \phi_{\text{l,k}} - U_{0,\text{k}}\left( c_{\text{s,k}}\rvert_{r_\text{k}=R_\text{k}} \right), \quad \mathrm{k} \in \{\mathrm{neg,\,pos}\},
\end{equation}
where $\phi_\text{s}$ is the potential in the solid phase and $U_0$ is the equilibrium potential, also known as open-circuit potential or open-circuit voltage (OCV), which depends on the surface concentration of the electrode particles.

\textit{Boundary and initial conditions:}
On the left and right boundary of the domain, no current flows out of the electrolyte, i.e., 
\begin{equation}
    i_\text{l,neg}\rvert_{x=0}=0 \quad \text{and} \quad i_\text{l,pos}\rvert_{x=L}=0,
\end{equation} 
where $L=L_\text{neg}+L_\text{sep}+L_\text{pos}$ is the sum of the length of each domain.

On the interface of negative electrode and separator, the potentials in the liquid phase are equal 
\begin{equation}
    \phi_\text{l,neg}\rvert_{x=L_\text{neg}}=\phi_\text{l,sep}\rvert_{x=L_\text{neg}} 
\end{equation}
and the currents are equal to the applied current
\begin{equation}
    i_\text{l,neg}\rvert_{x=L_\text{neg}}=i_\text{l,sep}\rvert_{x=L_\text{neg}}=I.
\end{equation}
The same holds for the interface of the separator and the positive electrode
\begin{equation}
\begin{split}
    \phi_\text{l,sep}\rvert_{x=L-L_\text{pos}}&=\phi_\text{l,pos}\rvert_{x=L-L_\text{pos}} \quad \text{and} \\
    i_\text{l,sep}\rvert_{x=L-L_\text{pos}}&=i_\text{l,pos}\rvert_{x=L-L_\text{pos}}=I.
\end{split}
\end{equation}

Akin to the currents, the molar fluxes in the electrolyte phase are zero at the left and right boundaries, i.e., 
\begin{equation}
    N_{\text{l,neg}}\rvert_{x=0}=0, \quad N_{\text{l,pos}}\rvert_{x=L}=0.
\end{equation}
And similarly, at both interfaces of the electrodes and the separator the concentrations in the liquid phase and the molar fluxes are equal, which means that at the interface of the negative electrode 
\begin{equation}
\begin{split}
    c_\text{l,neg}\rvert_{x=L_\text{neg}}&=c_\text{l,sep}\rvert_{x=L_\text{neg}} \quad \text{and} \\ N_\text{l,neg}\rvert_{x=L_\text{neg}}&=N_\text{l,sep}\rvert_{x=L_\text{neg}},
\end{split}
\end{equation} 
and at the interface of the positive electrode we have 
\begin{equation}
\begin{split}
    c_\text{l,sep}\rvert_{x=L-L_\text{pos}}&=c_\text{l,pos}\rvert_{x=L-L_\text{pos}} \quad \text{and} \\ N_\text{l,sep}\rvert_{x=L-L_\text{pos}}&=N_\text{l,pos}\rvert_{x=L-L_\text{pos}}.
\end{split}
\end{equation}

Inside the electrode particles, a zero Dirichlet condition is enforced at the center of the particle for the molar flux 
\begin{equation}
    N_{\text{s,k}}\rvert_{r_\text{k}=0}=0, \quad \text{k}\in\{\mathrm{neg,\,pos}\},
\end{equation}
while at the particle surface, the molar flux depends on the faradaic current density, such that 
\begin{equation}
    N_{\text{s,k}}\rvert_{r_\text{k}=R_\text{k}}=\frac{j_\text{k}}{F}, \quad \text{k}\in\{\mathrm{neg,\,pos}\}.
\end{equation}

Finally, we set initial values as initial conditions for the concentrations in the liquid and solid phases, i.e., 
\begin{equation}
\begin{split}
    c_\text{s,k}(x,r,t=0)&=c_{\text{s,k},0}, \quad \text{k}\in\{\mathrm{neg,\,pos}\} \quad \text{and} \\ c_\text{l,k}(x,t=0)&=c_{\text{l,k},0}, \quad \text{k}\in\{\mathrm{neg,\,sep,\,pos}\}.
\end{split}
\end{equation}

In the following, the parameters of the model are considered scalars independent of state of charge or concentration. Since we consider the isothermal case and do not include a thermal model, the model parameters are also not dependent on temperature.

\section{Methods}

\subsection{General}
For the numerical treatment of the equations described in section \ref{sec:DFN}, we rely on the DFN model implementation provided by the Python-based open-source framework \texttt{PyBaMM}~\cite{sulzer_python_2021}. We use \texttt{PyBaMM} v24.9.0. The base parametrization of our model is derived from the parameter set by Marquis et al.~\cite{marquis_asymptotic_2019} for a Kokam SLPB78205130H cell. Since we sample uncertain parameter vectors, the base parametrization only provides values for the parameters not varied over the course of the sensitivity analysis, e.g., the OCV curves, the electrode area, etc. We employ the CasADi solver~\cite{andersson_casadi_2018} and the default discretization, i.e., 20 mesh nodes in each subdomain in the $x$-dimension, as well as 20 mesh nodes in the $r$-dimension of each particle.

Since the parameter space is relatively vast, some randomly drawn parameter vectors might lead to convergence issues for the solver. We remedy this by omitting any parameter vectors that lead to non-converged simulations. This is well-justified because we utilize an independent and identically distributed random sampling approach and not quasi-random samples or quadrature-based approaches~\cite{sheikholeslami_what_2019}. The number of samples reported in section \ref{sec:results} pertains to the total number of simulation runs irrespective of convergence. However, the observed number of failed simulations does not exceed $0.003$\% for $N=100000$, which is negligibly small.

We employ a composite trapezoidal rule as the numerical quadrature rule mentioned in section \ref{sec:alexanderian}. All simulations and computations were conducted on a Fujitsu Celsius workstation with two Intel Xeon Gold 6128 CPUs and 128~GB of RAM. Shared-memory parallelization of our code is achieved using \texttt{joblib}~\cite{joblibDocumentation}. Our Python implementation of the algorithms outlined in section \ref{sec:alexanderian} and used throughout this work is available on GitHub\footnote{Source code available from \url{https://github.com/ezonta/TDGSA}}.

\subsection{Implementational details}
\label{sec:implementation}
The approaches described in section \ref{sec:alexanderian} depend heavily on generalized PC surrogate models to either approximate the model response or the KL modes. As mentioned above, these PC expansions are exact in the limit of infinite terms, but need to be truncated for reasons of computational feasibility. The classical truncation scheme retains only polynomials with total degree up to some integer value $p$, such that $||\boldsymbol{\alpha}||_1 \leq p$, where $\boldsymbol{\alpha}$ is the vector of exponents of a multivariate polynomial in the expansion. We call $p$ the \textit{order} of the PC expansion and consider expansions truncated this way \textit{full} expansions. Since this truncation scheme leads to $N_{\text{PC}} = \frac{(d + p)!}{d!p!}$ polynomials and corresponding coefficients, this ansatz quickly becomes impractical for high-dimensional problems with many uncertain parameters, at least for higher order PC expansions~\cite{sudret_global_2008}. We therefore also explore the use of an alternative truncation scheme better suited for higher dimensional cases based on the $q$-quasi-norm, namely \textit{hyperbolic cross truncation}, which specifies
\begin{equation}
    ||\boldsymbol{\alpha}||_q \leq p, \quad \text{where} \quad ||\boldsymbol{\alpha}||_q \coloneqq \left( \sum_{i=1}^d \alpha_i^q \right)^{\frac{1}{q}} 
\end{equation}
and thereby removes a lot of high-order interactions for $q\in(0,1)$, while keeping main effects and low-order interactions in the expansion~\cite{blatman_adaptive_2011}. The rationale behind preferably eliminating high-order interactions lies in the \textit{sparsity-of-effects principle}~\cite{montgomery_design_2013}. We refer to the resulting expansions as \textit{sparse} PC expansions. We employ \texttt{chaospy}~\cite{feinberg_chaospy_2015} for PC expansions and related probabilistic tools.

To determine the coefficients of the PC expansion and obtain a surrogate model, we use linear regression. Specifically, we utilize both ordinary least squares regression (OLS) and least angle regression (LARS) and compare their results in section \ref{sec:results}. While OLS targets the titular \textit{least squares problem}, i.e., the minimization of the squared residual, LARS aims at minimizing the $L_1$ norm. This makes LARS particularly suited for high-dimensional regression problems with limited data, especially since LARS is able to further sparsify the PC expansion by removing polynomials it deems irrelevant to the regression target. LARS achieves this by setting all coefficients to zero and increasing them one after another based on their correlation to the data. However, it also comes with the drawback of shrinkage on the coefficient estimates, which means that they are biased towards zero~\cite{efron_least_2004}. We rely on \texttt{scikit-learn}~\cite{pedregosa_2011} for its implementation of both regression algorithms.

We observe that in some cases the eigenvalues obtained from the discretized covariance operator deviate from the total variance estimated by the PC surrogate model of the KL mode. Our implementation of the algorithm based on KL expansions therefore checks for a relative error smaller than $10$\% between the variance derived from the eigenvalues and the PC coefficients before computing the Sobol' indices. If the error exceeds this bound, the sum of PC variances is used instead of the sum of dominant eigenvalues in the denominator of (\ref{eq:KL-sobol}). The mismatch does not seem to stem from an insufficiently sampled covariance matrix and the adjustment produces Sobol' indices in good agreement with those obtained via (\ref{eq:PC-sobol}). The cause of this divergence will have to be investigated further.

\subsection{Study design}
\subsubsection{Considered parameters and parameter ranges}

In the present study, we consider 24 parameters of the DFN model, namely the maximum solid phase concentration $c_{\text{s}}^\text{max}$, reaction rate constant $k_0$, electrical conductivity $\sigma$, solid phase diffusivity $D_\text{s}$, and particle radius $R$ of the positive and negative electrode; the layer thickness $L$, porosity $\varepsilon_\text{l}$, and Bruggemann coefficient $b$ of the positive electrode, separator, and negative electrode; as well as the thermodynamic factor $\Theta$, liquid phase diffusivity $D_\text{l}$, initial electrolyte concentration $c_{\text{l},0}$, ionic conductivity of the electrolyte $\kappa$, and the transference number $t^+$.

\begin{figure*}[tbph]
    \centering
    \includegraphics[width=\textwidth]{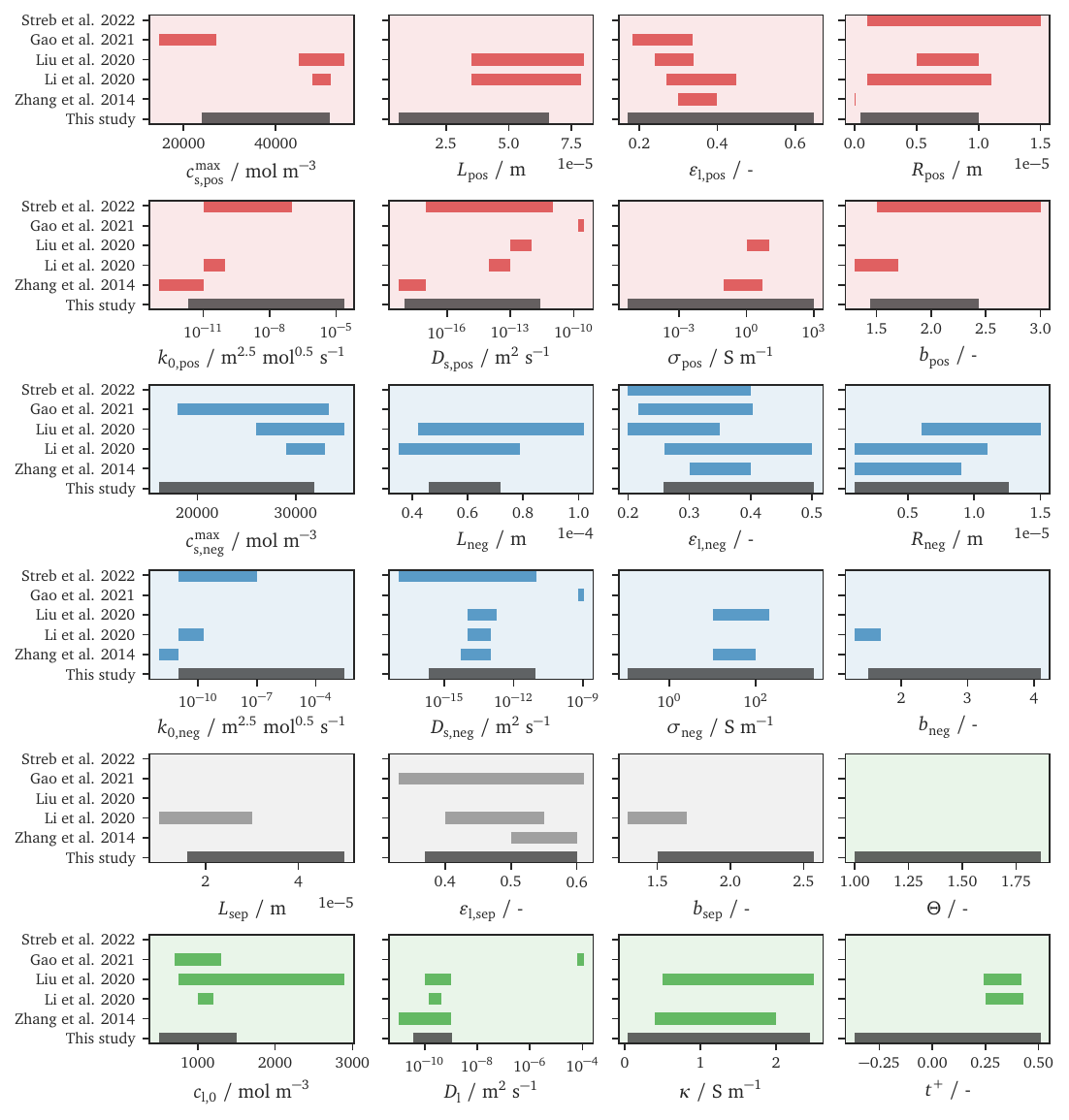}
    \caption{An overview of the parameter ranges covered in this study and in similar works by Zhang et al.~\cite{zhang_parameter_2014}, Li et al.~\cite{li_parameter_2020}, Liu et al.~\cite{liu_simulation_2020}, Gao et al.~\cite{gao_global_2021}, and Streb et al.~\cite{streb_improving_2022}. The parameter ranges obtained from the \texttt{LiionDB} agree well with most of the literature ranges, albeit with a tendency of being more extensive. However, some parameter intervals are smaller than their counterparts from literature. For $L_\text{neg}$, this can be explained by the availability of only two data points. The discrepancy between our value range for $c_{\text{l},0}$ and the interval of Liu et al. is relativized by the fact that various literature sources were utilized to construct the range. The obtained interval for $t^+$ extends into the negative numbers, which is not considered in any of the selected literature.}
    \label{fig:param_range}
\end{figure*}

In order to obtain sensible parameter ranges that reflect the actual prevalence in battery modeling research, we utilized the \texttt{LiionDB} database devised by Wang et al.~\cite{wang_parameter_2024}. The \texttt{LiionDB} collects battery model parameters reported in the literature. We queried the database for data from lithium nickel manganese cobalt oxide (NMC) positive electrodes regardless of composition, graphite negative electrodes, and LiPF$_6$-based electrolytes. We took into account all available data points at different temperatures, states of charge, etc., to further generalize the intervals. There was no data available in the database for $c_{\text{l},0}$, which is why we resorted to basing its range off of values from literature~\cite{edouard_parameter_2016,li_parameter_2020,wang_parameter_2024,park_optimal_2018,gao_global_2021}. The validity of negative transference numbers is debated~\cite{gouverneur_negative_2018, harris_comment_2018}. Nevertheless, we opted to keep the obtained intervals from the \texttt{LiionDB} unchanged. The parameter ranges used in this work and ranges from comparable studies are shown in Fig.~\ref{fig:param_range}. Numerical values can be found in Tab.~\ref{tab:param-ranges}.

The electrode area $A$ is kept constant throughout the study to get consistent current densities. We do not sample the solid volume fraction $\varepsilon_\text{s}$, as it would be dependent on $\varepsilon_\text{l}$, which would break the independency assumption of the methodology and would require more sophisticated treatment~\cite{kucherenko_estimation_2012}. For reasons of simplicity, we set $\varepsilon_\text{s} = 1 - \varepsilon_\text{l}$ and do not account for an inactive material volume fraction, as we do not suspect its influence to extend beyond the caused change in cell capacity. As is typical for sensitivity analyses, we sample scalar parameters and do not consider dependence of parameters on state of charge or concentration, because sampling functional parameters requires significantly more effort~\cite{iooss_global_2009}. We also do not sample the initial solid concentration $c_{\text{s},0}$ for each electrode, but rather set it to $\frac{1}{2}c_{\text{s}}^\text{max}$, such that the initial voltage is the same for every sampled parameter vector, because
\begin{equation}
    U(t=0) = U_{0,\text{pos}}\left(\frac{c_{\text{s},0,\text{pos}}}{c_{\text{s,pos}}^\text{max}}\right) - U_{0,\text{neg}}\left(\frac{c_{\text{s},0,\text{neg}}}{c_{\text{s,neg}}^\text{max}}\right).
\end{equation}
This of course leads to a randomized cell balancing, which will be addressed in section \ref{sec:current}.

\subsubsection{Current profiles} \label{sec:current}
Since each sampled parameter vector will lead to a different cell capacity, the applied current profile will have to be appropriate for the whole range of resulting capacities. The actual capacity, however, can not easily be obtained from the sampled parameter values, due to the randomized balancing. This means that we need some estimate for the cell capacity and that we ought to tailor our current magnitude such that we do not exceed too large C-rates in the low capacity samples, which might lead to convergence issues. To that end, we compute the theoretical limiting areal capacity $C_\text{theo}$ of each parameter vector
\begin{equation}
    C_\text{theo} = \min \left( C_{\text{theo,neg}}, C_{\text{theo,pos}} \right),
\end{equation}
where 
\begin{equation}
    C_\text{theo,k} = F \cdot c_\text{s,k}^\text{max} \cdot L_\text{k} \cdot \varepsilon_\text{l,k}, \quad \text{k}\in\{\mathrm{neg,\,pos}\}.
\end{equation}
As $C_\text{theo}$ is an estimate of the cell capacity neglecting electrode utilization determined by the cell balancing, it admits a theoretical $C_\text{theo}$-rate by dividing the applied current density through $C_\text{theo}$. This $C_\text{theo}$-rate is a lower bound estimate of the actual $C$-rate, coinciding in the case of full utilization of the limiting electrode, and will be used in the following to contrive the current magnitudes of the applied current profiles. In the case of underutilization of the limiting electrode in the considered voltage range, the actual $C$-rate will be higher than the $C_\text{theo}$-rate.

\begin{figure*}[hbt]
    \centering
    \includegraphics[width=\textwidth]{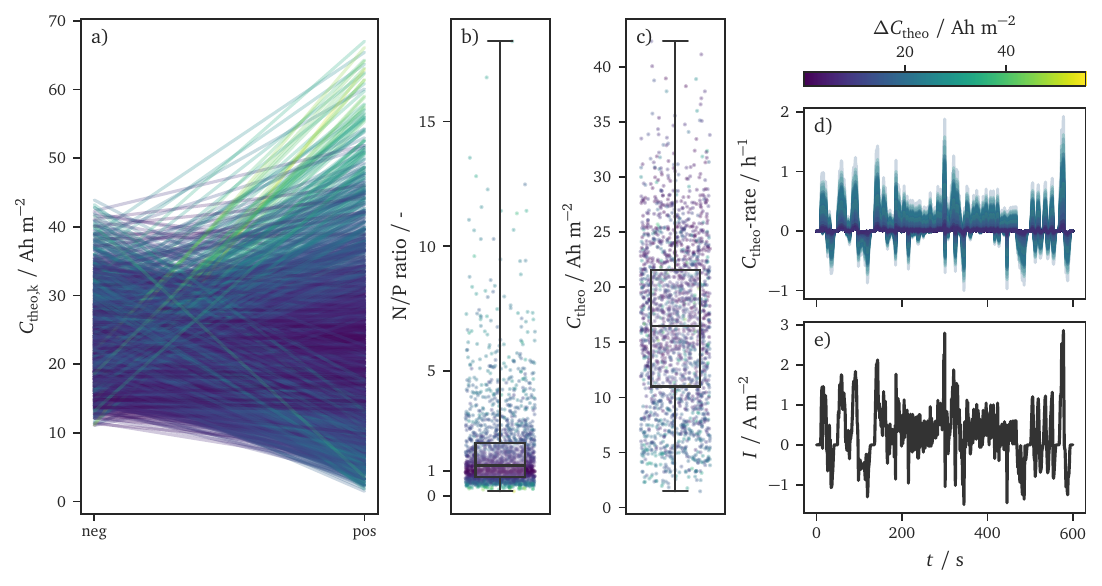}
    \caption{Visualization of the theoretical capacity of 2000 representative parameter samples to outline the scaling procedure for the drive cycle load profile. a) The balancing of the parameter samples depicted as a spaghetti plot connecting the theoretical capacity of the negative electrode with the theoretical capacity of the positive electrode. b) The resulting N/P ratio, i.e., the ratio of the capacity of the negative electrode over the capacity of the positive electrode. We can see that the median value is slightly larger than one, which agrees well with typical cells. c) A scatter plot of the limiting theoretical capacity, which corresponds to the minimum of each line in the spaghetti plot. In around 60\% of cases, the positive electrode is the limiting one. Since the spread is quite wide, the current magnitude of the drive cycle needs to be scaled, such that most simulations converge. In case of full utilization of the limiting electrode, the theoretical limiting capacity is the capacity of the cell. The scaled drive cycle is presented as d) the theoretical $C$-rates related to the theoretical limiting capacities and e) as a current density. The color of the lines and dots indicates the absolute difference in electrode capacity.}
    \label{fig:current-profiles}
\end{figure*}

We investigate the simulation of a highly dynamic current profile in this study, namely a scaled version of the US06 drive cycle, delivered with \texttt{PyBaMM} up until v24.1. The scaling is chosen such that the peak $C_\text{theo}$-rate does not exceed 2~h$^{-1}$. We could arguably have chosen a larger scaling factor at the expense of a potentially higher number of non-converged simulations, but the error investigation for the unscaled drive cycle seems to suggest that the results can to some extent be generalized for different scaling factors (Section \ref{sec:error}). Fig.~\ref{fig:current-profiles} visualizes and summarizes our deliberations on the relationship of theoretical limiting capacity and current magnitude for the load profile. The large variation in N/P-ratios (Fig.~\ref{fig:current-profiles}b) indicates that the depth of discharge will vary significantly between samples, making it difficult to draw conclusions with regard to the influence of the state of charge.

\subsubsection{Subgroup analysis}
We additionally perform a subgroup sensitivity analysis, aiming at a better resolved picture of the overall parametric sensitivity by decoupling high and low sensitivity scales. The grouping separates the parameters into parameters related to the negative electrode, positive electrode, separator, and electrolyte. In the negative and positive electrodes, the parameters are further subdivided into capacity-related and non-capacity-related parameters. This yields six groups in total, which are shown in Tab.~\ref{tab:group_params}. The capacity-related parameters of the positive and negative electrode are found in group 1 and 3 respectively and the non-capacity-related parameters of each electrode in group 2 and 4 correspondingly. The parameters related to the separator are in group 5 and the electrolyte-related parameters comprise group 6.

\begin{table}[htbp]
    \centering
    \caption{Parameter groupings for the subgroup sensitivity analysis based on domain affiliation.}
    \begin{tabular}{@{}ll@{}}
    \toprule
         Group & Parameters  \\ \midrule
         1 & $c_\text{s,pos}^\text{max}$, $L_\text{pos}$, $\varepsilon_\text{l,pos}$ \\
         2 & $b_\text{pos}$, $k_{0,\text{pos}}$, $\sigma_\text{pos}$, $D_\text{s,pos}$, $R_\text{pos}$ \\
         3 & $c_\text{s,neg}^\text{max}$, $L_\text{neg}$, $\varepsilon_\text{l,neg}$ \\
         4 & $b_\text{neg}$, $k_{0,\text{neg}}$, $\sigma_\text{neg}$, $D_\text{s,neg}$, $R_\text{neg}$ \\
         5 & $L_\text{sep}$, $\varepsilon_\text{l,sep}$, $b_\text{sep}$ \\
         6 & $\Theta$, $c_{\text{l},0}$, $D_\text{l}$, $\kappa$, $t^+$ \\
    \bottomrule
    \end{tabular}
    \label{tab:group_params}
\end{table}

In order to still obtain a sufficient exploration of parameter space in the subgroup analysis, we vary the fixed out-of-group parameters over a space-filling design of 30 points obtained from Latin hypercube sampling (LHS) over the whole parameter space. LHS attempts to cover the whole parameter space by sampling from a partition of equiprobable hypercubes, thereby stratifying the probability distribution~\cite{hadigol_least_2018}. We average the obtained Sobol' indices of each group over all these individual sensitivity analyses for different out-of-group parameter values.

\section{Results and discussion}
\label{sec:results}
We would like to preface this section with a short consideration: While the purely mathematical interpretation of Sobol' indices is rather straightforward, ascribing the sensitivity to some underlying physical processes described by the model is not as trivial and requires substantial domain knowledge. The matter is further complicated by the influence of the parameter ranges and current profile on the analysis. A \textit{general} sensitivity analysis, in the sense that it assesses the parametric sensitivity of the DFN model regardless of use case is therefore unfeasible. In the following, we aim to provide a discussion on general takeaways both methodologically and with regard to the DFN model, with the latter tending to be a bit more speculative, since sensitivity analysis does not provide any evidence on the underlying model behavior.

\subsection{Full analysis} \label{sec:full}
\begin{figure}[hptb]
    \centering
    \includegraphics[width=0.5\textwidth]{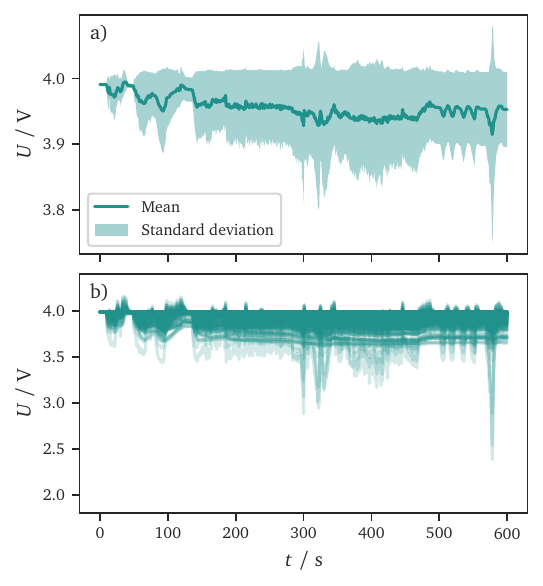}
    \caption{Output of the DFN model for the scaled US06 drive cycle for different parameter samples depicted as a) the mean and standard deviation over 10000 samples and b) individual voltage responses for 500 samples.}
    \label{fig:outputs-high-dynamics}
\end{figure}

Fig.~\ref{fig:outputs-high-dynamics} shows the voltage response over time of the simulations conducted to determine the generalized Sobol' indices when applying the scaled US06 drive cycle current profile. Most of the voltage responses seem to lie between $3.5-4.0$~V, while some samples reach significantly lower voltages, a few even being deeply discharged to voltages of around $2$~V for short time spans (Fig.~\ref{fig:outputs-high-dynamics}b). In the case of a real cell, discharging below the lower cutoff voltage of around $2.5$~V, depending on cell chemistry, may cause irreversible damage~\cite{qian_abuse_2016}. However, since the DFN model does not account for any of these degradation mechanisms, cell behavior is not altered by low cell voltages.

\begin{figure}[hptb]
    \centering
    \includegraphics[width=0.5\textwidth]{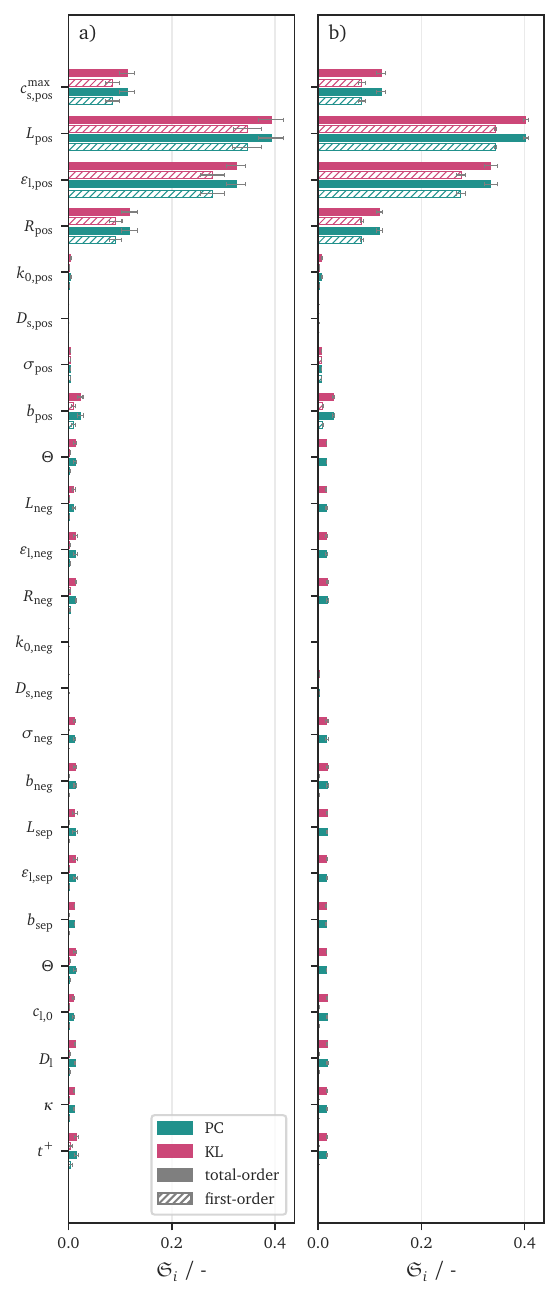}
    \caption{Generalized Sobol' indices obtained via three runs of sensitivity analysis with two different PC expansion truncation strategies, namely a) a full PC expansion of order 2 with $10000$ samples, $N_\text{KL}=10$, and OLS regression; and b) a sparse PC expansion of order 5 with $q=0.7$, $70000$ samples, $N_\text{KL}=10$, and OLS regression. The error bars signify the standard deviation of three individual runs.}
    \label{fig:sobol-high-dynamics}
\end{figure}

Based on the variance caused in the voltage response by the randomly sampled parameter vectors, we can obtain generalized Sobol' indices that aggregate the effects of each parameter on the model output over the whole time interval. The results of the complete sensitivity analysis over all considered parameters with a full and a sparse PC expansion are shown in Fig.~\ref{fig:sobol-high-dynamics}. We show the results obtained via OLS regression, because we did not find any significant differences between LARS and OLS (Fig.~\ref{fig:sobol_num_samples_LARS_OLS}). Unexpectedly, even for smaller sample sizes, LARS did not perform significantly better than OLS. We find that the capacity-determining parameters of the cathode, i.e., $c_{\text{s,pos}}^{\max}$, $L_\text{pos}$, $\varepsilon_{\text{l,pos}}$, together with the positive electrode particle radius $R_\text{pos}$ are responsible for almost all the variance in the voltage response, overshadowing the sensitivity of the remaining parameters. $L_\text{pos}$ is hereby the most influential parameter, closely followed by $\varepsilon_{\text{l,pos}}$, while $R_\text{pos}$ and $c_{\text{s,pos}}^{\max}$ each only amount to less than half the sensitivity of the more sensitive parameters. The difference in first-order and total-order indices is generally smaller than $0.1$, indicating a limited role of constructive --- i.e., purely variance increasing and non-compensating --- parameter interactions on the voltage variance. While it does seem as if some of the obfuscated insensitive parameters have total-order indices close to zero, suggesting that these are even more insignificant than those with slightly visible total-order indices, the size and uniformity of the small total-order indices as well as the practical nonexistence of corresponding first-order indices challenges that assumption. It is therefore necessary to consult the subgroup analysis (Section \ref{sec:subgroup}) to further resolve the influence of the lowly sensitive parameters. 

The predominant generalized Sobol' indices of these positive electrode parameters lie within expectation for multiple reasons: Firstly, the capacity-related parameters of the limiting electrode directly influence the full cell capacity and thereby the rate of change of the voltage when applying a current. Since the positive electrode is the limiting electrode in over 60\% of samples (Fig.~\ref{fig:current-profiles}), this means that the positive electrode has a higher influence on the overall cell capacity and therefore a higher influence on the cell voltage. This capacity-induced change in the voltage of the cell tends to be much higher than those from overpotentials due to transport limitations or similar effects. Moreover, the OCV of the positive electrode extends over a larger voltage window and has a tendentially steeper slope. Since the overall cell voltage is the difference between the potential of the positive and negative electrode, the voltage will change more drastically when the positive electrode is the limiting electrode. Finally, the influence of $R_\text{pos}$, which is wholly unrelated to the capacity of the cell, most probably stems from its relation to the specific surface area of the electrode. In the DFN model, the specific surface area $a_\text{pos}$ is computed via
\begin{equation}
    a_\text{pos} = 3 \frac{\varepsilon_\text{s,pos}}{R_\text{pos}}
\end{equation}
and governs the current density at the electrode-electrolyte interface. Increasing the current density leads to higher overpotentials in the electrode, influencing the cell voltage. Seemingly, this effect is much more pronounced in the positive electrode, hinting at a tendency of lower overpotentials in the negative electrode. The particle radius additionally determines the diffusion path length of the intercalated lithium, which presents another mode of influence on the overpotential. 

While the above does not constitute substantially novel insight into the DFN model, the results are interesting methodologically. First and foremost, they show that battery dynamics apparently produce low-rank processes, which lend themselves to sensitivity analysis via spectral representations. This is deducible from the close agreement of the generalized Sobol' indices obtained from both methods in Fig.~\ref{fig:sobol-high-dynamics} as well as the spectral decay of the eigenvalues, which is not shown here. This is an enormous advantage, because the KL method is much more efficient than the PC method. For our problem, where each of the 601 time steps is needed as a quadrature node due to insufficient smoothness of the output, the KL method needs to compute and store only about $1.7$\% of the number of coefficients used in the PC method. This is because the required number of surrogate models is reduced from 601 to 10. Depending on the number of cores available, this reduction can lead to solving the regression problem for all PC surrogate models in parallel, thereby significantly speeding up the procedure. This enables the use of higher order PC expansions for the KL method, since the number of coefficients is the limiting factor memory-wise. We observe that full lower order expansions yield comparable results to sparse higher order expansions for the generalized Sobol' indices. Higher-order PC expansions typically lead to more expressive surrogate models and lower surrogate model errors. Minimizing the surrogate model error is seemingly not immediately relevant for the analysis, but a low error might prove valuable for subsequent use cases of the surrogate model (Section \ref{sec:conclusion}). By comparison of the full (Fig.~\ref{fig:sobol-high-dynamics}a) and the sparse PC expansion (Fig.~\ref{fig:sobol-high-dynamics}b), we can also see that the choice of the hyperbolic cross truncation coefficient did not overly eliminate any relevant interaction terms in the expansion, since the total-order indices closely agree between the full and the sparse expansion, further substantiating that the sparsity-of-effects principle holds true.

We obtain estimates of the optimal number of samples (\ref{app2}) for the full second order PC expansion of $N=7475$ and $N=68379$ for the sparse PC expansion of order 5 with a cross truncation coefficient of $q=0.7$. This reveals that both analyses are provided with a sufficient number of samples. The error bars, however, do differ significantly between the analysis conducted with a sparse PC expansion and the analysis done with a full PC expansion. This likely stems from the much higher sample volume per run in the sparse case.

\subsection{Subgroup analysis} \label{sec:subgroup}

\begin{figure}[htbp]
    \centering
    \includegraphics[width=0.5\textwidth]{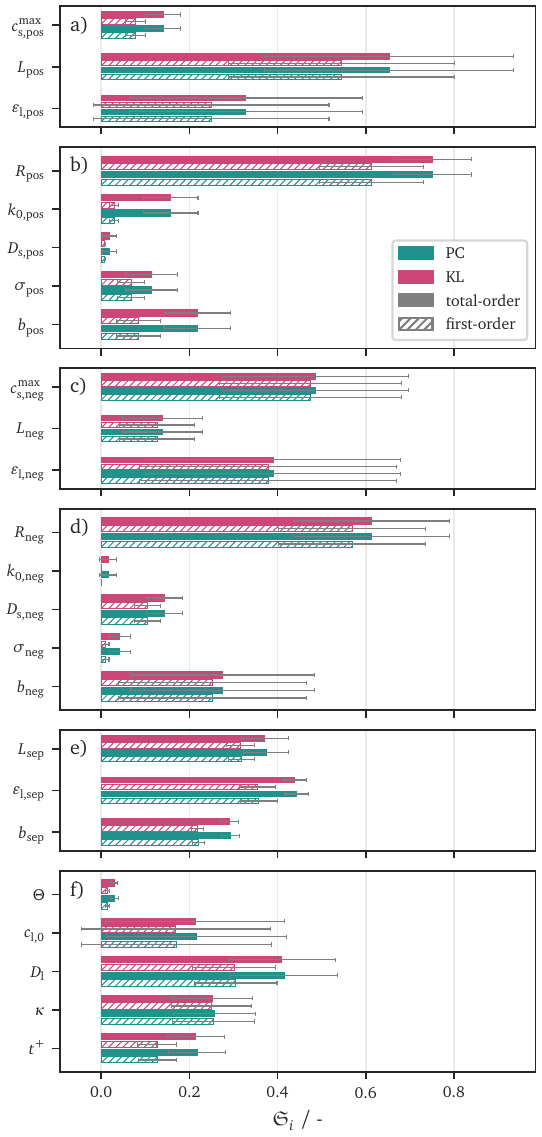}
    \caption{Results of the subgroup analysis for the six considered parameter groups, namely a) the capacity-related parameters of the positive electrode, b) the non-capacity-related parameters thereof, c) the capacity-related parameters of the negative electrode, d) the non-capacity-related parameters thereof, e) the seperator-related parameters, and f) the electrolyte-related parameters. Each sensitivity analysis was conducted using a full PC expansion of order 6 with 10000 samples, $N_\text{KL}=10$, and OLS regression. The error bars indicate the spread of the generalized Sobol' indices over the 30 different out-of-group parameter variations, which span over a design of space-filling LHS points.}
    \label{fig:grouped-high-dynamics}
\end{figure}

\begin{figure}
    \centering
    \includegraphics[width=0.5\textwidth]{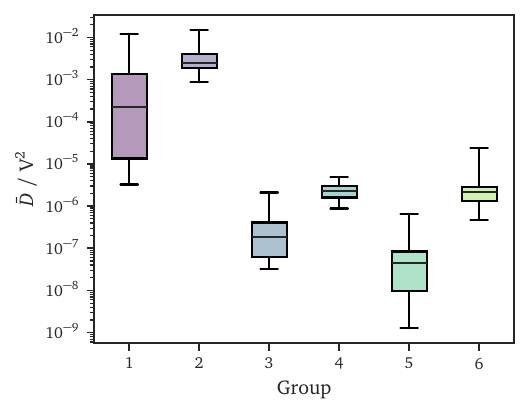}
    \caption{Boxplots of the mean caused variance in the voltage response per group over the 30 conducted sensitivity analyses with 10000 parameter samples. The mean caused variance $\bar{D}$ is the average of the variance over the whole time window of the model response. The group numbers refer to the groups in Tab.~\ref{tab:group_params}, where 1 stands for the capacity-related parameters of the positive electrode, 2 stands for the non-capacity-related parameters of the positive electrode, 3 and 4 are the capacity-related and non-capacity-related parameters of the negative electrode respectively, 5 includes the separator-related parameters, and 6 contains the electrolyte-related parameters.} 
    \label{fig:caused-variance}
\end{figure}

Fig.~\ref{fig:grouped-high-dynamics} shows the mean generalized Sobol' indices resulting from the subgroup sensitivity analysis where out-of-group parameters were varied over a space-filling design of LHS points. The error bars signify the variability of the indices when conducting the analysis at different points in parameter space. Since the scales are no longer coupled by a common total variance, we can not directly compare the magnitude of the indices between groups. For that reason, we have to take into consideration the mean caused variance per group $\bar{D}$, which offers a semi-quantitative hierarchical ordering of the groups (Fig.~\ref{fig:caused-variance}). We can see that the results of the subgroup analysis, together with the maximum caused variance, align well with the full analysis. The group of capacity-related positive electrode parameters causes the largest variance in the voltage response, akin to the full analysis, where most of the total variance is attributed to those same parameters. Similarly, the only parameter not related to the capacity with significant influence in the full analysis $R_\text{pos}$ is found to be the most important member in the group with the second highest variance. These observations suggest that the subgroup analysis does not yield contrasting results to the full analysis. Since the differences between first- and total-order indices in the subgroups tend to overall be rather small, and we assume inter-group interactivity to be much less relevant than intra-group interactivity due to our domain-based grouping, we believe that the interactive terms lost by subgrouping are negligible. Using the information provided by Fig.~\ref{fig:sobol-high-dynamics}, Fig.~\ref{fig:grouped-high-dynamics}, and Fig.~\ref{fig:caused-variance} lets us better resolve the coarse relative importance ranking provided by the full analysis.

Some of the indices in Fig.~\ref{fig:grouped-high-dynamics} come with high standard deviations, sometimes even higher than their mean value, e.g., in the case of $c_{\text{l},0}$ and $\sigma_\text{neg}$. This variability stems from the variation in the fixed out-of-group parameters and indicates changes in the physical behavior of the DFN model depending on the location in parameter space. An example of this could be $b_\text{neg}$, which controls the tortuosity of the negative electrode and is therefore highly important when lithium ion transport through the liquid phase in the porous negative electrode is the main limitation, but less relevant when other effects are limiting. In a full analysis, all of those cases are taken into account automatically, but in the subgroup analysis this is only emulated by averaging over different locations in parameter space. Nevertheless, the subgroup analysis is a very efficient way of complementing the full analysis to obtain more highly resolved hierarchical information.

Special consideration has to be given to the interpretation of the sensitivity indices of the capacity-related parameters. Akin to the full analysis, the positive electrode is capacity-limiting in most cases. Specifically, out of all the 30 analyses of group 1, approximately two thirds were dominated by a majority, i.e., over 50\%, of samples with a capacitively smaller positive electrode. The spread to lower variance values of group 1 in Fig.~\ref{fig:caused-variance} can likely be attributed to the other third, in which the positive electrode parameters are less influential by virtue of not being capacitively limiting. Group 3, i.e., the group of capacity-related parameters of the negative electrode, has significantly lower caused variance values compared to group 1. This is explainable by the lower number of analyses of group 3 with a majority of samples with a capacitively smaller negative electrode as well as the lower OCV range in the negative electrode, causing a smaller voltage change even when the negative electrode is capacitively limiting. Additionally, the relative importances in group 3 are presumably slightly skewed by the potentially insufficient range of $L_\text{neg}$ obtained from the \texttt{LiionDB} (Fig.~\ref{fig:param_range}). These aspects, together with the fact that, while less relevant for the voltage variance, the capacity of the non-limiting electrode is still highly relevant for the OCV balancing of the model and should not be unduly neglected, make the interpretation of variance-based sensitivity analysis complicated for capacity-related parameters. A potential workaround could be scaling the current profile for each parameter sample individually, thereby eliminating the influence of the capacity on the voltage response. This would also rid the dominance of capacity effects from the full analysis, potentially eliminating the need for a subgroup analysis. Future research could be directed at the feasibility and implications of a sensitivity analysis with individually scaled load profiles.

Furthermore, some general trends appear to emerge from the results. It seems as if in the negative electrode, charge transfer limitation rarely occurs, since the reaction rate constant, which drives the exchange current density, has barely any impact in its group. Conversely, in the positive electrode, the reaction rate constant does play a role, but the effect of the solid diffusivity seems negligible, indicating a subordinate role of solid diffusion limitation. This would likely be different for load profiles with longer consecutive charging or discharging times or with larger current densities because charge transfer or solid diffusion limitations would have an overall bigger impact on the dynamics of the cell.

\subsection{Parameter fixing and error estimation} \label{sec:error}

Alexanderian et al.~\cite{alexanderian_2020} discuss the role of the generalized Sobol' indices as a priori error bounds for the loss of total variance when fixing unimportant parameters to some nominal values inside the corresponding bounds. From the viewpoint of battery model parametrization, we are less interested in the loss of variance of the stochastic model and would rather know about the error introduced to a parametrized battery model with deterministic parameter values when setting the unimportant parameters to some random value inside their bounds. This is to emulate the common practice in battery modelling of setting some parameters to arbitrary values from literature, and investigate its influence on the model error. To this end, we employ the base parametrization to simulate the unscaled version of the US06 drive cycle and vary increasingly more insensitive parameters to examine the resulting model error. The model error in this case is the deviation from the reference result obtained with the base parametrization. The seven most insensitive parameters determined above consist of $b_\text{sep}$, $L_\text{sep}$, $\varepsilon_\text{l,sep}$, $k_{0,\text{neg}}$, $\sigma_\text{neg}$, $\Theta$, and $t^+$. In the following, the insensitive parameters are always varied over 100 samples.

\begin{figure*}[hbtp]
    \centering
    \includegraphics[width=\textwidth]{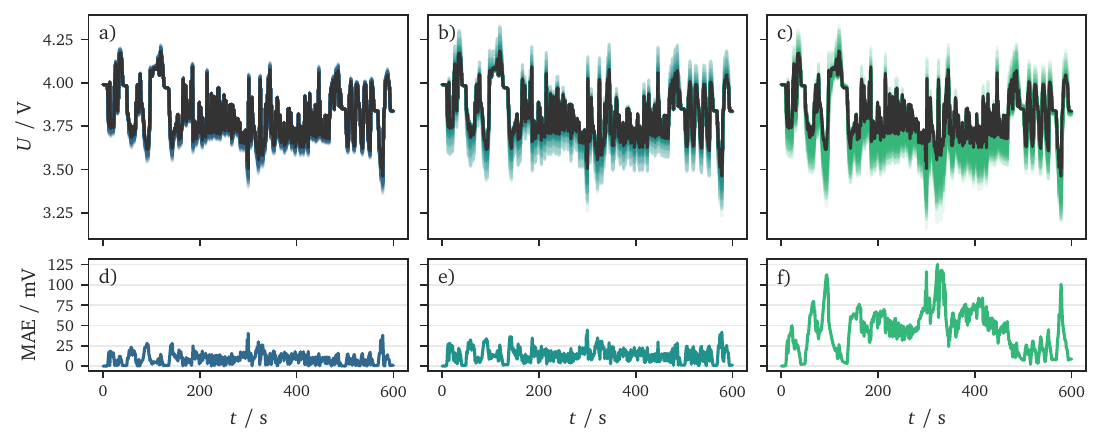}
    \caption{Comparison of the voltage responses obtained via the base parametrization and when varying an increasing amount of insensitive parameters. The voltage curve comparison for a) three ($b_\text{sep}$, $L_\text{sep}$, $\varepsilon_\text{l,sep}$), b) five ($b_\text{sep}$, $L_\text{sep}$, $\varepsilon_\text{l,sep}$, $\sigma_\text{neg}$, $k_{0,\text{neg}}$), and c) seven ($b_\text{sep}$, $L_\text{sep}$, $\varepsilon_\text{l,sep}$, $\sigma_\text{neg}$, $k_{0,\text{neg}}$, $\Theta$, $t^+$) insensitive parameters as well as the corresponding mean absolute error plots for variations in the d) three, e) five, and f) seven least influential parameters substantiate the results of the subgroup analysis.}
    \label{fig:insensitive}
\end{figure*}

Fig.~\ref{fig:insensitive} shows the results of selecting random values for an increasing number of insensitive parameters and comparing the voltage response to the reference output. When varying the three least sensitive parameters, i.e., $b_\text{sep}$, $L_\text{sep}$, and $\varepsilon_\text{l,sep}$, the mean absolute error (MAE) mostly stays below $25$~mV (Fig.~\ref{fig:insensitive}d) and the average root mean squared error (RMSE) over all simulations is $11.2$~mV. For random variations of the five least sensitive parameters, the MAE largely stays around $25$~mV (Fig.~\ref{fig:insensitive}e) and the average RMSE amounts to $17.0$~mV. Setting the seven least important parameters to random values inside their respective intervals causes an MAE above $100$~mV (Fig.~\ref{fig:insensitive}f) and an average RMSE of $53.6$~mV.

These results underscore the low impact of changes in insensitive parameters on the model response and show that the methodology outlined above yields valid sensitivity information over the whole time interval. However, it also exemplifies a shortcoming: sensitivity information is always related to a simulative scenario, a certain load profile, etc., and is also influenced by our choice of the distribution for the parameter values. This hinders general statements on relative parametric importance, because one might be able to devise a load profile which leads to wholly different parametric sensitivities by design. However, since models are usually employed for certain use cases and only parametrized and validated for these use cases, one should generally be able to investigate the parametric sensitivity for select representative scenarios of battery model use (e.g., fast charging, discharging, and a drive cycle for use in the automotive sector) to get a good understanding of which parameters are less influential in the region of model validity. Instead of using very general uniform parameter bounds, one might also choose normally distributed parameters taking into account some experimental error. This can answer the question which parameter's error contributes most to the battery model's uncertainty, in order to potentially remeasure said parameter to decrease the error and improve the model. The generalized investigation of the input-output relations of the DFN model is of course also limited by the differences in cell chemistry. While a cell with an NMC positive electrode and a graphite negative electrode exhibits an OCV with a significant slope for every state of charge, a LiFePO$_4$ positive electrode typically plateaus in medium ranges of the state of charge. This affects the model behavior in numerous ways and is difficult to consider concomitantly in a sensitivity study.

Finally, we would generally advise against considering capacity-related parameters as insensitive from the above study. As aforementioned, the interplay of the capacity related parameters with the electrode and total cell capacity is complicated and not immediately visible in a large voltage variance. Nevertheless, the quality of the model highly depends on correct capacitive balancing, which stems from the combination of the capacity-related parameters of both electrodes. To alleviate this problem in potential future studies with broad parameter ranges, one could be conducting sensitivity studies with individually scaled load profiles, to nullify the effect of capacity change on the voltage response and thereby yield capacity-agnostic sensitivities.

\section{Conclusion and outlook}
\label{sec:conclusion}
Efficient and accurate sensitivity analysis of nonlinear electrochemical battery models is of high relevance for a variety of use cases and research questions. This work restated the problem regarding the inadequacy of OAT methods, discussed the present restriction of readily available global sensitivity analysis methods to scalar quantities of interest, and implemented novel approaches by Alexanderian et al.~\cite{alexanderian_2020} for global sensitivity analysis of time-dependent model outputs. We applied the methods to the simulation of a drive cycle with the DFN model and determined generalized Sobol' indices by a full analysis and a subgroup analysis to further resolve the influence of lowly sensitive parameters. We found that battery models seemingly produce low rank processes, which lend themselves to efficient time-dependent global sensitivity analysis via spectral representations. Specifically, the KL method produced comparable results to the PC method at approximately one hundredth of the memory requirements. The computational cost can be lowered even further by the use of sparse PC expansions. In cases where the low rank process requirement might not be met, the PC method would be the preferred choice. Moreover, subsequent simulations with random values inserted for an increasing number of unimportant parameters revealed a direct correlation to the model error, corroborating the sensitivity analysis results and suggesting potential applications in battery model parametrization.

Unfortunately, sensitivity indices are dependent on the chosen load profile and the selected parameter distributions, which limits the generalizability of the results and makes it hard to derive universal input-output relations of the DFN model. However, we were able to determine certain pitfalls in variance-based sensitivity analysis of electrochemical battery models, with the hope of providing some guidance in employing this method as a helpful tool in the field of simulative battery research.

Overall, this work presents an important step towards efficient and accurate sensitivity analyses for time-dependent quantities in simulation-based battery research and beyond. The presented methodology enables researchers to determine model parameters of special interest for factor prioritization via generalized first-order Sobol' indices, and less influential parameters for factor fixing via generalized total-order Sobol' indices. Research questions concerned with the main factors driving some model behavior over time can therefore be addressed more easily and adequately. This is especially relevant for model-based design studies, where the parameters of physical systems are optimized using well-validated simulation models, because the most influential parameters can be identified more reliably. One such example would be model-based cell design using the DFN model.

Further research could be directed towards the feasibility of capacity-agnostic sensitivity analyses of electrochemical battery models with individually scaled load profiles. More sophisticated models, e.g., the DFN model coupled to a thermal model to account for heat generation and temperature dependence, should be investigated in future sensitivity studies to be able to more realistically describe cell behaviour. Extending the approach to be able to account for functional dependencies of parameters would be a worthwhile feature in this context. Additionally, the created surrogate models with a low enough approximation error might be employed in downstream tasks where cheap function evaluations are required, e.g., optimization, potentially with a multifidelity ansatz. Moreover, one could investigate alternative implementational choices, e.g., a different quadrature rule for the treatment of the time integrals in the computation of the generalized Sobol' indices, different optimization algorithms for the determination of the PC coefficients or employing the pseudo-spectral approach for computing PC coefficients.

\section*{Declaration of competing interest}

Michele Spinola and Christoph Weißinger are employees at Capgemini Engineering. The remaining authors declare that the research was conducted in the absence of any commercial or financial relationships that could be construed as a potential conflict of interest.

\section*{CRediT authorship contribution statement}

\textbf{Elia Zonta:} Conceptualization, Methodology (lead), Software, Formal analysis (equal), Investigation, Data curation, Writing - original draft, Visualization.
\textbf{Ivana Jovanovic Buha:} Methodology (supporting), Formal analysis (equal), Writing - review \& editing.
\textbf{Michele Spinola:} Writing - review \& editing, Project administration.
\textbf{Christoph Weißinger:} Writing - review \& editing, Supervision, Funding acquisition.
\textbf{Hans-Joachim Bungartz:} Writing - review \& editing, Supervision.
\textbf{Andreas Jossen:} Writing - review \& editing, Supervision, Funding acquisition.

\section*{Acknowledgements}
This work is supported by Capgemini in the context of the TUM-Capgemini Research and Development Agreement ``Innovative and non-invasive methods of parameter identification for lithium-ion battery models''. The authors would like to thank Johannes Natterer, Alexander Frank, and Axel Durdel for fruitful discussions.

\appendix

\section{Considered parameter ranges}
\label{app1}

Tab.~\ref{tab:param-ranges} summarizes the parameter distributions that were used to conduct sensitivity analysis throughout this work.

\begin{table*}[htbp]
    \centering
    \caption{Numerical values of the parameter ranges used throughout this work.}
    \begin{tabular}{@{}llllll@{}}
    \toprule
       Name & Symbol & Unit & Min. & Max. & Distribution  \\  \midrule 
       Positive electrode max, concentration & $c_\text{s,pos}^\text{max}$ & mol m$^{-3}$ & 23900 & 51765 & Uniform \\
       Positive electrode thickness & $L_\text{pos}$ & m & $6.00\cdot 10^{-6}$ & $6.60\cdot 10^{-5}$ & Uniform \\
       Positive electrode porosity & $\varepsilon_\text{l,pos}$ & - & 0.171 & 0.648 & Uniform \\
       Positive electrode particle radius & $R_\text{pos}$ & m & $5.00\cdot 10^{-7}$ & $1.00\cdot 10^{-5}$ & Uniform \\
       Positive electrode reaction rate constant & $k_{0,\text{pos}}$ & m$^{2.5}$ mol$^{-0.5}$ s$^{-1}$ & $2.10\cdot 10^{-12}$ & $2.41\cdot 10^{-5}$ & Loguniform \\
       Positive electrode diffusivity & $D_\text{s,pos}$ & m$^2$ s$^{-1}$ & $9.59\cdot 10^{-19}$ & $2.51\cdot 10^{-12}$ & Loguniform \\
       Positive electrode conductivity & $\sigma_\text{pos}$ & S m$^{-1}$ & $5.20\cdot 10^{-6}$ & $1.00\cdot 10^{3}$ & Loguniform \\
       Positive electrode Bruggeman coefficient & $b_\text{pos}$ & - & $1.44$ & $2.44$ & Uniform \\
       Negative electrode max, concentration & $c_\text{s,neg}^\text{max}$ & mol m$^{-3}$ & 16100 & 31920 & Uniform \\
       Negative electrode thickness & $L_\text{neg}$ & m & $4.60\cdot 10^{-5}$ & $7.20\cdot 10^{-5}$ & Uniform \\
       Negative electrode porosity & $\varepsilon_\text{l,neg}$ & - & $0.26$ & $0.50$ & Uniform \\
       Negative electrode particle radius & $R_\text{neg}$ & m & $1.00\cdot 10^{-5}$ & $1.26\cdot 10^{-5}$ & Uniform \\
       Negative electrode reaction rate constant & $k_{0,\text{neg}}$ & m$^{2.5}$ mol$^{-0.5}$ s$^{-1}$ & $1.00\cdot 10^{-11}$ & $3.00\cdot 10^{-3}$ & Loguniform \\
       Negative electrode diffusivity & $D_\text{s,neg}$ & m$^2$ s$^{-1}$ & $2.00\cdot 10^{-16}$ & $9.07\cdot 10^{-12}$ & Loguniform \\
       Negative electrode conductivity & $\sigma_\text{neg}$ & S m$^{-1}$ & $1.11\cdot 10^{-1}$ & $2.20\cdot 10^{3}$ & Loguniform \\
       Negative electrode Bruggeman coefficient & $b_\text{neg}$ & - & $1.50$ & $4.10$ & Uniform \\
       Separator thickness & $L_\text{sep}$ & m & $1.60\cdot 10^{-5}$ & $5.00\cdot 10^{-5}$ & Uniform \\
       Separator porosity & $\varepsilon_\text{l,sep}$ & - & $0.37$ & $0.60$ & Uniform \\
       Separator Bruggeman coefficient & $b_\text{sep}$ & - & $1.50$ & $2.57$ & Uniform \\
       Thermodynamic factor & $\Theta$ & - & $1.00$ & $1.86$ & Uniform \\
       Initial electrolyte concentration & $c_{\text{l},0}$ & mol m$^{-3}$ & $500$ & $1500$ & Uniform \\
       Electrolyte diffusivity & $D_\text{l}$ & m$^2$ s$^{-1}$ & $3.60\cdot 10^{-11}$ & $1.09\cdot 10^{-9}$ & Loguniform \\
       Electrolyte ionic conductivity & $\kappa$ & S m$^{-1}$ & $4.45\cdot 10^{-3}$ & $2.45$ & Uniform \\
       Transference number & $t^+$ & - & $-0.37$ & $0.51$ & Uniform \\
    \bottomrule
    \end{tabular}
    \label{tab:param-ranges}
\end{table*}

\section{Influence of sample size and choice of regression algorithm}
\label{app2}

\begin{figure*}[p]
    \centering
    \includegraphics[width=\textwidth]{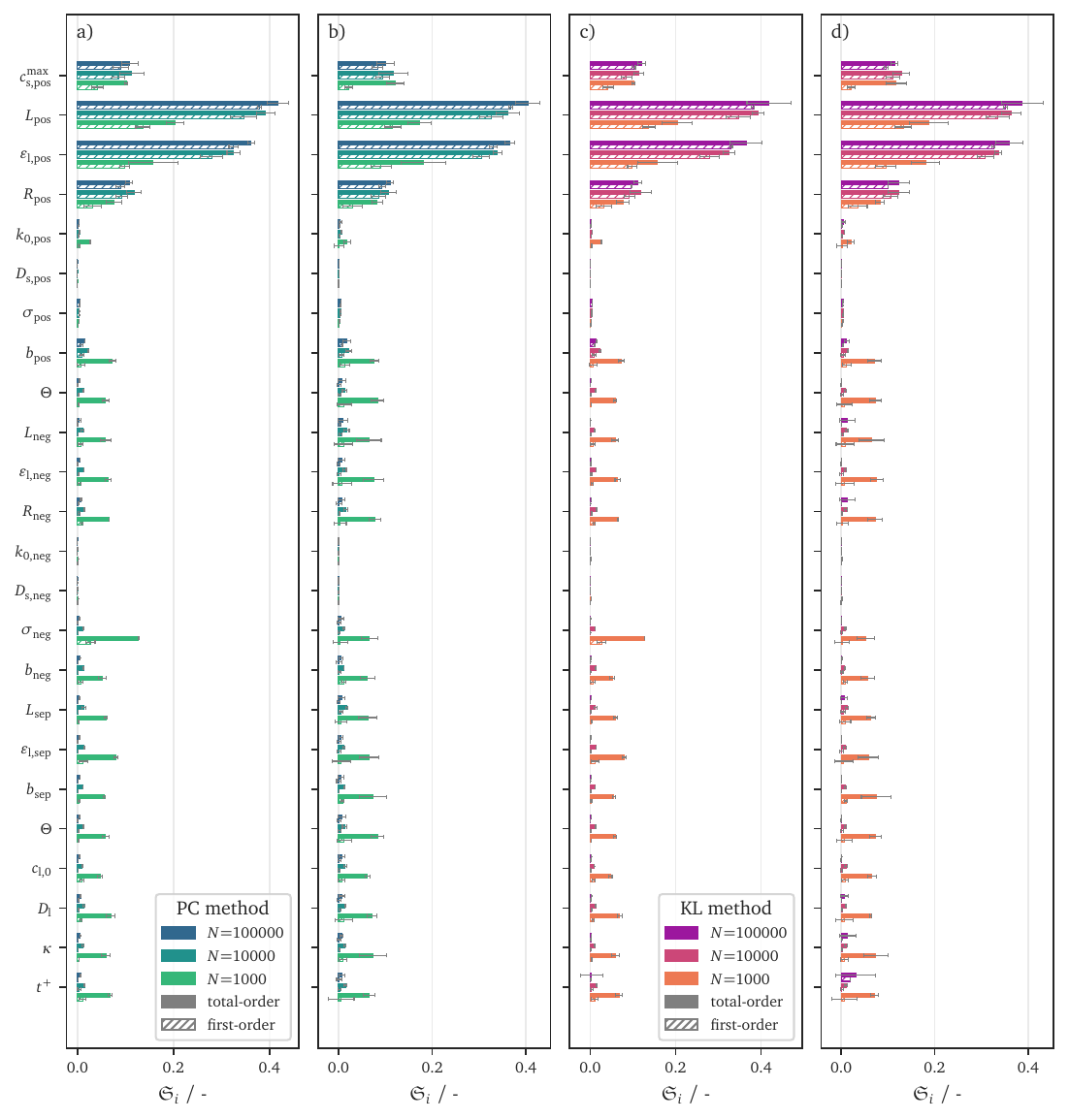}
    \caption{Results of different sensitivity analyses with a full PC expansion of order 2, varying sample sizes, $N_\text{KL}=10$, using a) the PC method with OLS regression, b) the PC method with LARS regression, c) the KL method with OLS regression, and d) the KL method with LARS regression.}
    \label{fig:sobol_num_samples_LARS_OLS}
\end{figure*}

Fig.~\ref{fig:sobol_num_samples_LARS_OLS} shows the results of a small robustness study concerning the time-dependent global sensitivity analysis. We investigated the influence of the number of samples and of the two different regression algorithms on the outcome of the sensitivity analysis over three individual runs each.

Sudret~\cite{sudret_global_2008} provides an empirical estimate for the optimal number of samples for PC-based surrogate models, given as
\begin{equation}
    N = (d-1)N_\text{PC}, 
\end{equation}
where the number of PC terms or coefficients can be computed for full expansions as
\begin{equation}
    N_\text{PC} = \frac{(d+p)!}{d!p!}.
\end{equation}
This lets us compute the optimal number of samples for the robustness study as $N=7475$. We can see that above this sample size threshold, the results vary only slightly, while the analysis with a sample size of $N=1000$ appears to be significantly undersampled.

The results obtained with OLS and LARS regression appear to be almost equivalent, irrespective of sample size.

\section{Comparison of one-at-a-time sensitivity analysis and generalized Sobol' indices}
\label{app3}
\begin{figure*}[hbtp]
    \centering
    \includegraphics[width=\textwidth]{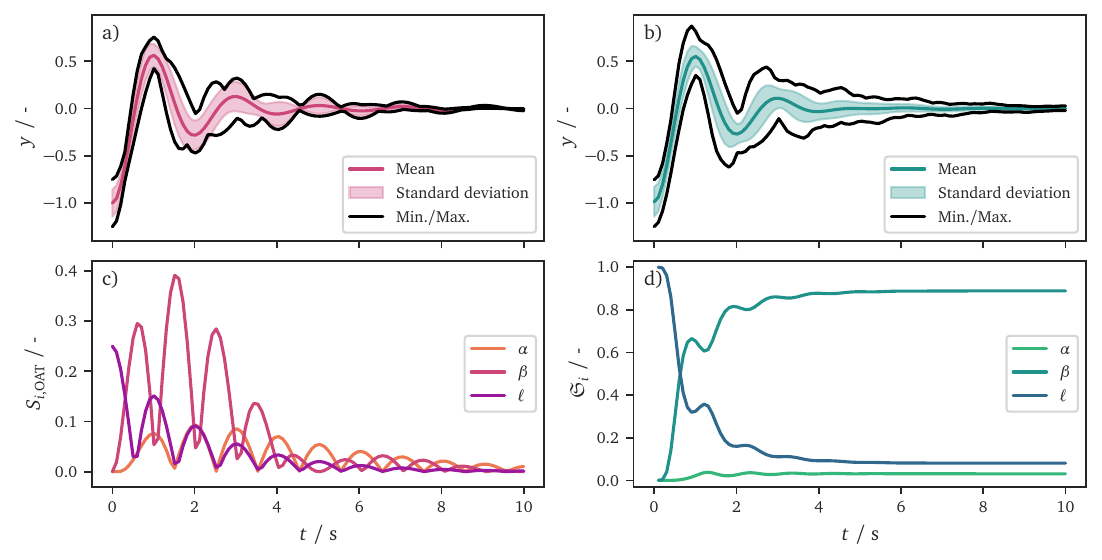}
    \caption{Comparison of OAT sensitivity with generalized Sobol' indices. The upper row shows the damped harmonic oscillator displacement obtained from a) varying each parameter individually to its minimum and maximum value and b) sampling 100 random parameter combinations. The lower row displays the corresponding c) OAT sensitivity metrics and d) generalized Sobol' indices.}\label{fig:OAT_vs_Sobol}
\end{figure*}

To better illustrate the theoretical considerations made in section \ref{sec:inadequacy}, we compare the result of one-at-a-time sensitivity analysis with the generalized Sobol' indices used throughout this work for a damped oscillator example from literature~\cite{alexanderian_2020}, whose displacement $y$ is governed by an initial value problem of the form
\begin{equation}
    \begin{split}
        &y'' + 2\alpha y' + (\alpha^2 + \beta^2)y = 0, \\
        &y(0) = \ell, \quad y'(0)=0,
    \end{split}
\end{equation}
where $\alpha\sim\mathcal{U}(\frac{3}{8},\frac{5}{8})$, $\beta\sim\mathcal{U}(\frac{10}{4},\frac{15}{4})$, $\ell\sim\mathcal{U}(-\frac{5}{4},-\frac{3}{4})$ are considered uncertain parameters with uniform distributions. For this comparison, we employ the OAT sensitivity metric from Li et al.~\cite{li_parameter_2020}, namely
\begin{equation}
    S_{i,\text{OAT}} = \sqrt{\frac{1}{N_s}\sum_{j=1}^{N_s}(y_{ij}-\bar{y}_i)^2},
\end{equation}
where $y_{ij}$ is the displacement when changing a parameter $i$ to a certain level $j$, $\bar{y}_i$ is the average over the displacements for all levels of a certain parameter, and $N_s$ is the number of levels. We set $N_s=2$ and use the bounds of the uniform distribution for each parameter. Since $y$ is a function of time, we obtain one $S_{i,\text{OAT}}$ per parameter for each point in time. Generalized Sobol' indices are obtained using the PC method using 100 quadrature nodes in time, PC expansions of order 4, and 100 random samples. Fig.~\ref{fig:OAT_vs_Sobol} compares the results of both methods.

The displacement obtained when varying each parameter individually to its minimum and maximum values shows that the attainable maximum and minimum displacements are underestimated (Fig.~\ref{fig:OAT_vs_Sobol}a). When parameters are varied in tandem, the displacement reaches lower minima and higher maxima, especially in the early phase (Fig.~\ref{fig:OAT_vs_Sobol}b). This means that the possible amplitude of the model response in the given parameter space is underestimated when parameters are only varied individually. Comparing the different sensitivity indices reveals that they show similar trends, which stems from the rather weak coupling of the parameters in the damped oscillator example. However, the pointwise-in-time $S_{i,\text{OAT}}$ values lack conclusive interpretability, because the ranking of parametric importance differs at each point in time. Contrastingly, generalized Sobol' indices at a certain time point $t$ aggregate the parametric effects of the whole interval $[0,t]$, making their interpretation straightforward and yielding a single metric describing the normalized sensitivity over the whole time interval.

For battery models, which generally exhibit much stronger nonlinear coupling between the parameters, the differences in model output observed in Fig.~\ref{fig:OAT_vs_Sobol}a and b will be exacerbated. The agreement in trend between OAT and generalized Sobol' indices will generally also decrease with increasing nonlinear coupling between model parameters.

Fig.~\ref{fig:OAT_vs_Sobol}d has the additional merit of showing that our implementation of the method correctly reproduces the original results of Alexanderian et al.~\cite[Fig.~2]{alexanderian_2020}.

\bibliographystyle{elsarticle-num} 
\bibliography{literature.bib}

\end{document}